\documentclass[%
reprint, 
superscriptaddress,
preprintnumbers,
nofootinbib,
nobibnotes,
 amsmath,amssymb,
 aps,
]{revtex4-2}
\setcitestyle{authoryear,round}
\usepackage[colorlinks=true]{hyperref}
\hypersetup{
    urlcolor=black,
    citecolor=blue,
    linkcolor=blue,
  }
\usepackage{graphicx}
\usepackage{dcolumn}
\usepackage{bm}
\usepackage[mathlines]{lineno}

\usepackage{physics}
\usepackage{cleveref}
\usepackage{comment}
\usepackage[dvipsnames]{xcolor}
\usepackage{xspace}
\usepackage{xcolor}
\usepackage[normalem]{ulem}

\usepackage{physics}
\usepackage{footnote}

\usepackage[T1]{fontenc}
\usepackage{aecompl}

\usepackage{mdframed,bm}
\usepackage[T1]{fontenc}
\usepackage{ae,aecompl}
\usepackage{amsmath}

\newcommand{\map}{\ensuremath{\left\langle M_{\mathrm{ap}}^3\right\rangle}}

\newcommand{\aap}{A\&A}
\newcommand{\apjs}{ApJS}
\newcommand{\aj}{A.J.}
\newcommand{\mnras}{MNRAS}
\newcommand{\physrep}{Phys. Rep.}

\newcommand{\jcap}{JCAP}
\newcommand{\apjl}{ApJ}
\newcommand{\pasj}{PASJ}

\begin{document}

\preprint{DES-2021-0681}
\preprint{FERMILAB-PUB-22-001-PPD-SCD}

\title{Dark Energy Survey Year 3 Results:\\Three-Point Shear Correlations and Mass Aperture Moments}

\author{L.~F.~Secco}\email{secco@uchicago.edu}
\affiliation{Kavli Institute for Cosmological Physics, University of Chicago, Chicago, IL 60637, USA}
\author{M.~Jarvis}
\affiliation{Department of Physics and Astronomy, University of Pennsylvania, Philadelphia, PA 19104, USA}
\author{B.~Jain}
\affiliation{Department of Physics and Astronomy, University of Pennsylvania, Philadelphia, PA 19104, USA}
\author{C.~Chang}
\affiliation{Kavli Institute for Cosmological Physics, University of Chicago, Chicago, IL 60637, USA}
\affiliation{Department of Astronomy and Astrophysics, University of Chicago, Chicago, IL 60637, USA}
\author{M.~Gatti}
\affiliation{Department of Physics and Astronomy, University of Pennsylvania, Philadelphia, PA 19104, USA}
\author{J.~Frieman}
\affiliation{Kavli Institute for Cosmological Physics, University of Chicago, Chicago, IL 60637, USA}
\affiliation{Department of Astronomy and Astrophysics, University of Chicago, Chicago, IL 60637, USA}
\affiliation{Fermi National Accelerator Laboratory, P. O. Box 500, Batavia, IL 60510, USA}
\author{S.~Adhikari}
\affiliation{Kavli Institute for Cosmological Physics, University of Chicago, Chicago, IL 60637, USA}
\affiliation{Department of Astronomy and Astrophysics, University of Chicago, Chicago, IL 60637, USA}
\author{\linebreak A.~Alarcon}
\affiliation{Argonne National Laboratory, 9700 South Cass Avenue, Lemont, IL 60439, USA}
\author{A.~Amon}
\affiliation{Kavli Institute for Cosmology, University of Cambridge, Madingley Road, Cambridge CB3 0HA, UK}
\affiliation{Institute of Astronomy, University of Cambridge, Madingley Road, Cambridge, CB3 0HA}
\author{K.~Bechtol}
\affiliation{Physics Department, 2320 Chamberlin Hall, University of Wisconsin-Madison, 1150 University Avenue Madison, WI  53706-1390}
\author{M.~R.~Becker}
\affiliation{Argonne National Laboratory, 9700 South Cass Avenue, Lemont, IL 60439, USA}
\author{G.~M.~Bernstein}
\affiliation{Department of Physics and Astronomy, University of Pennsylvania, Philadelphia, PA 19104, USA}
\author{J.~Blazek}
\affiliation{Department of Physics, Northeastern University, Boston, MA 02115, USA}
\affiliation{Laboratory of Astrophysics, \'Ecole Polytechnique F\'ed\'erale de Lausanne (EPFL), Observatoire de Sauverny, 1290 Versoix, Switzerland}
\author{A.~Campos}
\affiliation{Department of Physics, Carnegie Mellon University, Pittsburgh, Pennsylvania 15312, USA}
\author{A.~Carnero~Rosell}
\affiliation{Instituto de Astrofisica de Canarias, E-38205 La Laguna, Tenerife, Spain}
\affiliation{Laborat\'orio Interinstitucional de e-Astronomia - LIneA, Rua Gal. Jos\'e Cristino 77, Rio de Janeiro, RJ - 20921-400, Brazil}
\affiliation{Universidad de La Laguna, Dpto. Astrofísica, E-38206 La Laguna, Tenerife, Spain}
\author{M.~Carrasco~Kind}
\affiliation{Center for Astrophysical Surveys, National Center for Supercomputing Applications, 1205 West Clark St., Urbana, IL 61801, USA}
\affiliation{Department of Astronomy, University of Illinois at Urbana-Champaign, 1002 W. Green Street, Urbana, IL 61801, USA}
\author{A.~Choi}
\affiliation{California Institute of Technology, 1200 East California Blvd, MC 249-17, Pasadena, CA 91125, USA}
\author{J.~Cordero}
\affiliation{Jodrell Bank Center for Astrophysics, School of Physics and Astronomy, University of Manchester, Oxford Road, Manchester, M13 9PL, UK}
\author{J.~DeRose}
\affiliation{Lawrence Berkeley National Laboratory, 1 Cyclotron Road, Berkeley, CA 94720, USA}
\author{S.~Dodelson}
\affiliation{Department of Physics, Carnegie Mellon University, Pittsburgh, Pennsylvania 15312, USA}
\affiliation{NSF AI Planning Institute for Physics of the Future, Carnegie Mellon University, Pittsburgh, PA 15213, USA}
\author{C.~Doux}
\affiliation{Department of Physics and Astronomy, University of Pennsylvania, Philadelphia, PA 19104, USA}
\author{A.~Drlica-Wagner}
\affiliation{Department of Astronomy and Astrophysics, University of Chicago, Chicago, IL 60637, USA}
\affiliation{Fermi National Accelerator Laboratory, P. O. Box 500, Batavia, IL 60510, USA}
\affiliation{Kavli Institute for Cosmological Physics, University of Chicago, Chicago, IL 60637, USA}
\author{S.~Everett}
\affiliation{Santa Cruz Institute for Particle Physics, Santa Cruz, CA 95064, USA}
\author{G.~Giannini}
\affiliation{Institut de F\'{\i}sica d'Altes Energies (IFAE), The Barcelona Institute of Science and Technology, Campus UAB, 08193 Bellaterra (Barcelona) Spain}
\author{D.~Gruen}
\affiliation{Faculty of Physics, Ludwig-Maximilians-Universit\"at, Scheinerstr. 1, 81679 Munich, Germany}
\author{R.~A.~Gruendl}
\affiliation{Center for Astrophysical Surveys, National Center for Supercomputing Applications, 1205 West Clark St., Urbana, IL 61801, USA}
\affiliation{Department of Astronomy, University of Illinois at Urbana-Champaign, 1002 W. Green Street, Urbana, IL 61801, USA}
\author{I.~Harrison}
\affiliation{Department of Physics, University of Oxford, Denys Wilkinson Building, Keble Road, Oxford OX1 3RH, UK}
\affiliation{Jodrell Bank Center for Astrophysics, School of Physics and Astronomy, University of Manchester, Oxford Road, Manchester, M13 9PL, UK}
\affiliation{School of Physics and Astronomy, Cardiff University, CF24 3AA, UK}
\author{W.~G.~Hartley}
\affiliation{Department of Astronomy, University of Geneva, ch. d'\'Ecogia 16, CH-1290 Versoix, Switzerland}
\author{K.~Herner}
\affiliation{Fermi National Accelerator Laboratory, P. O. Box 500, Batavia, IL 60510, USA}
\author{E.~Krause}
\affiliation{Department of Astronomy/Steward Observatory, University of Arizona, 933 North Cherry Avenue, Tucson, AZ 85721-0065, USA}
\author{N.~MacCrann}
\affiliation{Department of Applied Mathematics and Theoretical Physics, University of Cambridge, Cambridge CB3 0WA, UK}
\author{J.~McCullough}
\affiliation{Kavli Institute for Particle Astrophysics \& Cosmology, P. O. Box 2450, Stanford University, Stanford, CA 94305, USA}
\author{J.~Myles}
\affiliation{Department of Physics, Stanford University, 382 Via Pueblo Mall, Stanford, CA 94305, USA}
\affiliation{Kavli Institute for Particle Astrophysics \& Cosmology, P. O. Box 2450, Stanford University, Stanford, CA 94305, USA}
\affiliation{SLAC National Accelerator Laboratory, Menlo Park, CA 94025, USA}
\author{A. Navarro-Alsina}
\affiliation{Instituto de F\'isica Gleb Wataghin, Universidade Estadual de Campinas, 13083-859, Campinas, SP, Brazil}
\author{J.~Prat}
\affiliation{Department of Astronomy and Astrophysics, University of Chicago, Chicago, IL 60637, USA}
\affiliation{Kavli Institute for Cosmological Physics, University of Chicago, Chicago, IL 60637, USA}
\author{R.~P.~Rollins}
\affiliation{Jodrell Bank Center for Astrophysics, School of Physics and Astronomy, University of Manchester, Oxford Road, Manchester, M13 9PL, UK}
\author{S.~Samuroff}
\affiliation{Department of Physics, Carnegie Mellon University, Pittsburgh, Pennsylvania 15312, USA}
\author{C.~S{\'a}nchez}
\affiliation{Department of Physics and Astronomy, University of Pennsylvania, Philadelphia, PA 19104, USA}
\author{I.~Sevilla-Noarbe}
\affiliation{Centro de Investigaciones Energ\'eticas, Medioambientales y Tecnol\'ogicas (CIEMAT), Madrid, Spain}
\author{E.~Sheldon}
\affiliation{Brookhaven National Laboratory, Bldg 510, Upton, NY 11973, USA}
\author{M.~A.~Troxel}
\affiliation{Department of Physics, Duke University Durham, NC 27708, USA}
\author{D.~Zeurcher}
\affiliation{Department of Physics, ETH Zurich, Wolfgang-Pauli-Strasse 16, CH-8093 Zurich, Switzerland}


\author{M.~Aguena}
\affiliation{Laborat\'orio Interinstitucional de e-Astronomia - LIneA, Rua Gal. Jos\'e Cristino 77, Rio de Janeiro, RJ - 20921-400, Brazil}
\author{F.~Andrade-Oliveira}
\affiliation{Department of Physics, University of Michigan, Ann Arbor, MI 48109, USA}
\author{J.~Annis}
\affiliation{Fermi National Accelerator Laboratory, P. O. Box 500, Batavia, IL 60510, USA}
\author{D.~Bacon}
\affiliation{Institute of Cosmology and Gravitation, University of Portsmouth, Portsmouth, PO1 3FX, UK}
\author{E.~Bertin}
\affiliation{CNRS, UMR 7095, Institut d'Astrophysique de Paris, F-75014, Paris, France}
\affiliation{Sorbonne Universit\'es, UPMC Univ Paris 06, UMR 7095, Institut d'Astrophysique de Paris, F-75014, Paris, France}
\author{S.~Bocquet}
\affiliation{Faculty of Physics, Ludwig-Maximilians-Universit\"at, Scheinerstr. 1, 81679 Munich, Germany}
\author{D.~Brooks}
\affiliation{Department of Physics \& Astronomy, University College London, Gower Street, London, WC1E 6BT, UK}
\author{D.~L.~Burke}
\affiliation{Kavli Institute for Particle Astrophysics \& Cosmology, P. O. Box 2450, Stanford University, Stanford, CA 94305, USA}
\affiliation{SLAC National Accelerator Laboratory, Menlo Park, CA 94025, USA}
\author{J.~Carretero}
\affiliation{Institut de F\'{\i}sica d'Altes Energies (IFAE), The Barcelona Institute of Science and Technology, Campus UAB, 08193 Bellaterra (Barcelona) Spain}
\author{F.~J.~Castander}
\affiliation{Institut d'Estudis Espacials de Catalunya (IEEC), 08034 Barcelona, Spain}
\affiliation{Institute of Space Sciences (ICE, CSIC),  Campus UAB, Carrer de Can Magrans, s/n,  08193 Barcelona, Spain}
\author{M.~Crocce}
\affiliation{Institut d'Estudis Espacials de Catalunya (IEEC), 08034 Barcelona, Spain}
\affiliation{Institute of Space Sciences (ICE, CSIC),  Campus UAB, Carrer de Can Magrans, s/n,  08193 Barcelona, Spain}
\author{L.~N.~da Costa}
\affiliation{Laborat\'orio Interinstitucional de e-Astronomia - LIneA, Rua Gal. Jos\'e Cristino 77, Rio de Janeiro, RJ - 20921-400, Brazil}
\affiliation{Observat\'orio Nacional, Rua Gal. Jos\'e Cristino 77, Rio de Janeiro, RJ - 20921-400, Brazil}
\author{M.~E.~S.~Pereira}
\affiliation{Department of Physics, University of Michigan, Ann Arbor, MI 48109, USA}
\affiliation{Hamburger Sternwarte, Universit\"{a}t Hamburg, Gojenbergsweg 112, 21029 Hamburg, Germany}
\author{J.~De~Vicente}
\affiliation{Centro de Investigaciones Energ\'eticas, Medioambientales y Tecnol\'ogicas (CIEMAT), Madrid, Spain}
\author{H.~T.~Diehl}
\affiliation{Fermi National Accelerator Laboratory, P. O. Box 500, Batavia, IL 60510, USA}
\author{P.~Doel}
\affiliation{Department of Physics \& Astronomy, University College London, Gower Street, London, WC1E 6BT, UK}
\author{K.~Eckert}
\affiliation{Department of Physics and Astronomy, University of Pennsylvania, Philadelphia, PA 19104, USA}
\author{I.~Ferrero}
\affiliation{Institute of Theoretical Astrophysics, University of Oslo. P.O. Box 1029 Blindern, NO-0315 Oslo, Norway}
\author{B.~Flaugher}
\affiliation{Fermi National Accelerator Laboratory, P. O. Box 500, Batavia, IL 60510, USA}
\author{D.~Friedel}
\affiliation{Center for Astrophysical Surveys, National Center for Supercomputing Applications, 1205 West Clark St., Urbana, IL 61801, USA}
\author{J.~Garc\'ia-Bellido}
\affiliation{Instituto de Fisica Teorica UAM/CSIC, Universidad Autonoma de Madrid, 28049 Madrid, Spain}
\author{G.~Gutierrez}
\affiliation{Fermi National Accelerator Laboratory, P. O. Box 500, Batavia, IL 60510, USA}
\author{S.~R.~Hinton}
\affiliation{School of Mathematics and Physics, University of Queensland,  Brisbane, QLD 4072, Australia}
\author{D.~L.~Hollowood}
\affiliation{Santa Cruz Institute for Particle Physics, Santa Cruz, CA 95064, USA}
\author{K.~Honscheid}
\affiliation{Center for Cosmology and Astro-Particle Physics, The Ohio State University, Columbus, OH 43210, USA}
\affiliation{Department of Physics, The Ohio State University, Columbus, OH 43210, USA}
\author{D.~Huterer}
\affiliation{Department of Physics, University of Michigan, Ann Arbor, MI 48109, USA}
\author{K.~Kuehn}
\affiliation{Australian Astronomical Optics, Macquarie University, North Ryde, NSW 2113, Australia}
\affiliation{Lowell Observatory, 1400 Mars Hill Rd, Flagstaff, AZ 86001, USA}
\author{N.~Kuropatkin}
\affiliation{Fermi National Accelerator Laboratory, P. O. Box 500, Batavia, IL 60510, USA}
\author{M.~A.~G.~Maia}
\affiliation{Laborat\'orio Interinstitucional de e-Astronomia - LIneA, Rua Gal. Jos\'e Cristino 77, Rio de Janeiro, RJ - 20921-400, Brazil}
\affiliation{Observat\'orio Nacional, Rua Gal. Jos\'e Cristino 77, Rio de Janeiro, RJ - 20921-400, Brazil}
\author{J.~L.~Marshall}
\affiliation{George P. and Cynthia Woods Mitchell Institute for Fundamental Physics and Astronomy, and Department of Physics and Astronomy, Texas A\&M University, College Station, TX 77843,  USA}
\author{F.~Menanteau}
\affiliation{Center for Astrophysical Surveys, National Center for Supercomputing Applications, 1205 West Clark St., Urbana, IL 61801, USA}
\affiliation{Department of Astronomy, University of Illinois at Urbana-Champaign, 1002 W. Green Street, Urbana, IL 61801, USA}
\author{R.~Miquel}
\affiliation{Instituci\'o Catalana de Recerca i Estudis Avan\c{c}ats, E-08010 Barcelona, Spain}
\affiliation{Institut de F\'{\i}sica d'Altes Energies (IFAE), The Barcelona Institute of Science and Technology, Campus UAB, 08193 Bellaterra (Barcelona) Spain}
\author{J.~J.~Mohr}
\affiliation{Faculty of Physics, Ludwig-Maximilians-Universit\"at, Scheinerstr. 1, 81679 Munich, Germany}
\affiliation{Max Planck Institute for Extraterrestrial Physics, Giessenbachstrasse, 85748 Garching, Germany}
\author{R.~Morgan}
\affiliation{Physics Department, 2320 Chamberlin Hall, University of Wisconsin-Madison, 1150 University Avenue Madison, WI  53706-1390}
\author{J.~Muir}
\affiliation{Perimeter Institute for Theoretical Physics, 31 Caroline St. North, Waterloo, ON N2L 2Y5, Canada}
\author{F.~Paz-Chinch\'{o}n}
\affiliation{Center for Astrophysical Surveys, National Center for Supercomputing Applications, 1205 West Clark St., Urbana, IL 61801, USA}
\affiliation{Institute of Astronomy, University of Cambridge, Madingley Road, Cambridge CB3 0HA, UK}
\author{A.~Pieres}
\affiliation{Laborat\'orio Interinstitucional de e-Astronomia - LIneA, Rua Gal. Jos\'e Cristino 77, Rio de Janeiro, RJ - 20921-400, Brazil}
\affiliation{Observat\'orio Nacional, Rua Gal. Jos\'e Cristino 77, Rio de Janeiro, RJ - 20921-400, Brazil}
\author{A.~A.~Plazas~Malag\'on}
\affiliation{Department of Astrophysical Sciences, Princeton University, Peyton Hall, Princeton, NJ 08544, USA}
\author{M.~Rodriguez-Monroy}
\affiliation{Centro de Investigaciones Energ\'eticas, Medioambientales y Tecnol\'ogicas (CIEMAT), Madrid, Spain}
\author{A.~Roodman}
\affiliation{Kavli Institute for Particle Astrophysics \& Cosmology, P. O. Box 2450, Stanford University, Stanford, CA 94305, USA}
\affiliation{SLAC National Accelerator Laboratory, Menlo Park, CA 94025, USA}
\author{E.~Sanchez}
\affiliation{Centro de Investigaciones Energ\'eticas, Medioambientales y Tecnol\'ogicas (CIEMAT), Madrid, Spain}
\author{S.~Serrano}
\affiliation{Institut d'Estudis Espacials de Catalunya (IEEC), 08034 Barcelona, Spain}
\affiliation{Institute of Space Sciences (ICE, CSIC),  Campus UAB, Carrer de Can Magrans, s/n,  08193 Barcelona, Spain}
\author{E.~Suchyta}
\affiliation{Computer Science and Mathematics Division, Oak Ridge National Laboratory, Oak Ridge, TN 37831}
\author{M.~E.~C.~Swanson}
\affiliation{National Center for Supercomputing Applications, 1205 West Clark St.,
Urbana, IL 61801, USA}
\author{G.~Tarle}
\affiliation{Department of Physics, University of Michigan, Ann Arbor, MI 48109, USA}
\author{D.~Thomas}
\affiliation{Institute of Cosmology and Gravitation, University of Portsmouth, Portsmouth, PO1 3FX, UK}
\author{C.~To}
\affiliation{Center for Cosmology and Astro-Particle Physics, The Ohio State University, Columbus, OH 43210, USA}
\author{J.~Weller}
\affiliation{Max Planck Institute for Extraterrestrial Physics, Giessenbachstrasse, 85748 Garching, Germany}
\affiliation{Universit\"ats-Sternwarte, Fakult\"at f\"ur Physik, Ludwig-Maximilians Universit\"at M\"unchen, Scheinerstr. 1, 81679 M\"unchen, Germany}

\collaboration{DES Collaboration}

\date{\today}

\begin{abstract}
We present high signal-to-noise measurements of three-point shear correlations and the third moment of the mass aperture statistic using the first 3 years of data from the Dark Energy Survey. We additionally obtain the first measurements of the configuration and scale dependence of the four three-point shear correlations which carry cosmological information. 
With the third-order mass aperture statistic, we present tomographic measurements over angular scales of 4 to 60 arcminutes with a combined statistical significance of 15.0$\sigma$.
Using the tomographic information and measuring also the second-order mass aperture, we additionally obtain a skewness parameter and its redshift evolution. We find that the amplitudes and scale-dependence of these shear 3pt functions are in qualitative agreement with measurements in a mock galaxy catalog based on N-body simulations, indicating promise for including them in future cosmological analyses. We validate our measurements by showing that B-modes, parity-violating contributions and PSF modeling uncertainties are negligible, and determine that the measured signals are likely to be of astrophysical and gravitational origin.    

\end{abstract}

\maketitle

\section{Introduction}\label{sec: intro}

 Two-point (2pt) auto-correlation functions of the shear field (sometimes referred to as cosmic shear) have been widely used in the recent literature to constrain cosmological parameters. Current works utilize different statistical measures and exploit the shear distributions in both real (configuration) space as well as harmonic space (\cite{hikage2019,asgari20,hamana20}, \citet{Amon21}, \citet*{Secco21}). One of the main products of years of effort by the community is the accurate determination of the amplitude parameter $S_8\equiv\sigma_8\sqrt{\Omega_\mathrm{m}/0.3}$, where $\sigma_8$ is the root mean square amplitude of the linear-theory matter power spectrum at $z=0$ over an 8 Mpc/$h$ scale, and $\Omega_\mathrm{m}$ is the matter density at $z=0$. This amplitude is in mild tension with the value inferred from fluctuations of the Cosmic Microwave Background \citep{Aghanim:2018eyx} by about 2$\sigma$ (depending on the survey data sample used) and its origin remains unresolved.

Extracting more cosmological information from the shear field than that encoded in 2pt statistics may help better characterize this tension and is an important goal in itself. To be useful, the additional information should have \textit{its systematics well-understood and controlled}. The aim of this work is to address both points above: we use data from DES Y3, the first 3 years of data from the Dark Energy Survey (\citet*{y3-gold, y3-shapecatalog,y3-3x2ptkp}) to obtain high signal-to-noise measurements of three-point (3pt) correlation functions of the shear field and show that potential contaminants in these measurements coming from observational and instrumental origins are negligible.

The benefits of utilizing higher order correlations as a cosmological probe are plenty and go far beyond simply enabling access to non-Gaussian information in the shear and matter bispectrum. Compared to 2pt functions, 3pt correlations in lensing carry different cosmological parameter degeneracies \citep{Takada_Jain_2002, Bernardeau2002,Kayo2013} and when combined with 2pt functions can additionally constrain astrophysical and systematic nuisance parameters \citep{Huterer2006, Troxel2012, Pyne2021, Semboloni2013}. The combination of 2pt and 3pt lensing data vectors is thus greater than the sum of its parts, and enables degeneracy-breaking in both the cosmological and nuisance parameter spaces. 

The community has followed several approaches to extracting the information contained in higher order shear statistics. For example, non-Gaussian information can be obtained with position-dependent or integrated 2pt lensing signatures \citep{Halder2021,Jung2021}, peak statistics \citep{Kacprzak2016,Zurcher2021}, density splits of the shear field \citep{Friedrich2018,Gruen2018} as well as with techniques borrowed from artificial intelligence and neural networks \citep{Fluri2019,Cheng2020,Jeffrey2021,Lu2021}. Another approach is to \textit{directly measure} 3rd or higher order statistics of the shear field in the form of ellipticity correlations (\citet{Waerbekeetal2002}, \cite{Benabed2006}), mass aperture moments \citep{JBJ04,Fu_etal_2014,Semboloni_etal_2011} or lensing mass maps \citep{Gatti2021moments}.

In this work, we follow the latter approach and directly measure 3pt statistics of the DES Y3 data in the form of ``natural'' correlation functions (the three-point equivalents of $\xi_\pm$) \citep{Schneider2003_natural_components} and the third moment of the mass aperture statistic  \citep{Schneider_etal_1998}.  We detect both statistics at high significance and additionally explore the triangle configuration dependence, tomographic signals and redshift evolution of the 3pt lensing signal, none of which have been previously measured at high significance in survey data. 

We also verify that several null tests of great importance for cosmological applications (such as B-mode contamination, PSF residual errors and parity-violating contributions) are consistent with zero or otherwise negligible compared to the $E$-mode signal for these 3pt statistics in DES Y3. This work, therefore, represents the first step towards a cosmological analysis with DES Y3 data using the statistics presented here, which we leave for the future.

This paper is structured as follows. In Sec. \ref{sec: data} we provide an overview of the DES Y3 weak lensing shear catalog  and an N-body simulation that we utilize as a check on the rough scale dependence and amplitude of the 3pt signatures. In Sec. \ref{sec: theory} we review the underlying theory of three-point lensing correlations as a probe of the matter bispectrum and describe the estimators we utilize in the data. In Sec. \ref{sec: results} we present the main results of this paper: the measured signals of the mass aperture skewness, natural shear correlations, and some explorations of their configuration and redshift dependence, as well as a comparison with existing detections. In Sec. \ref{sec: validation} we validate the measured signals and verify that their origin must be astrophysical and gravitational by checking that $B$-mode, PSF and parity-violating contaminations are negligible and that our data estimator is robust. We conclude and mention future avenues and challenges in Sec. \ref{sec: conclusion}.

\section{Data}\label{sec: data}

We describe below the data utilized in this work, the DES Y3 shape catalog and a simulated (N-body) mock. We regard the latter as providing a simplified theory estimate, serving as a basic check of the data measurement.

\subsection{DES Y3 Data}\label{sec: des data}

The first 3 years of data from the Dark Energy Survey (DES Y3) cover the full footprint of the survey's six-year campaign. Its nominal area is over 5,000 deg$^2$, which is reduced to 4143 deg$^2$ after data selections and cuts that optimize the observed samples for weak lensing and galaxy clustering measurements, with a baseline mask described in \cite{y3-gold}. The DES data were collected using the 570 megapixel Dark Energy Camera (DECam; \citet{flaugher15}) in five photometric bands $grizY$ at the Blanco telescope at Cerro Tololo Inter-American Observatory (CTIO) in Chile.

Here we are interested in the \textsc{Metacalibration} \citep{Sheldon_Huff_2017, Huff_Mandelbaum_2017} shape catalog produced and validated in the DES Y3 analysis  (\citet*{Gatti2021}). This is the largest shear catalog to date in number of objects and area, with over 100 million objects with a mean redshift of $z=0.63$ and a weighted source number density  $n_{\rm eff} = 5.59 \; \mathrm{arcmin}^{-2}$. An overview of the DES Y3 weak lensing and galaxy clustering cosmological
analysis is available in~\cite{y3-3x2ptkp},
where further specifications of the data and analysis tools are available in references
contained within.

In the DES Y3 cosmological analysis, source galaxies were separated into four redshift bins each with approximately equal numbers of galaxies (\citet*{Myles_Alarcon_2021}). In some of the measurements presented in this work, we also separate the shear data into tomographic bins. However, since the 3pt statistics have lower signal-to-noise than the 2pt measurements, we instead divide the DES Y3 \textsc{Metacalibration} catalog into just 2 redshift bins, which we label $z_1$ and $z_2$. The lower redshift bin, $z_1$, is a combination of the galaxies assigned to bins 1 and 2 in the fiducial analysis, while bin $z_2$ is a combination of the galaxies originally assigned to bins 3 and 4 in that analysis. Weighting the galaxy redshifts in these two newly defined bins by their inverse-variance ellipticity and shear response, we obtain mean redshifts $\left\langle z_1 \right\rangle=0.42$ and $\left\langle z_2 \right\rangle=0.81$ with widths of 0.30 and 0.27 respectively.

Since the shape catalog used to derive the cosmic shear results in DES Y3 (\cite{Amon21}, \citet*{Secco21}) has been extensively validated, we use the same data quality cuts and sample specification in the 3pt analysis below.

\subsection{T17 Mock Catalog}\label{sec: takahashi}

To support our findings reported in the following sections, the same 3rd order correlation measurement pipelines applied to DES Y3 data are also applied to an N-body mock galaxy catalog based on \cite{Takahashi2017} [hereafter T17].

We use full-sky lensing convergence and shear maps from T17 to create a DES Y3-like, tomographic shape catalog. In particular, we used a single one out of their 108 available sets of convergence and shear map snapshots, which span a redshift range between z = 0.05 and 5.3 at intervals of 150 $h^{ - 1}$ Mpc comoving distance. The maps have been obtained via ray-tracing using the algorithm \textsc{GRayTrix} \citep{Hamana2015}, based on the output of different N-body simulations. The N-body simulations have been run using the code \textsc{L-Gadget-2} \citep{springel05}, assuming a flat $\Lambda$CDM WMAP 9 cosmology \citep{Hinshaw2013} with parameters given by $(\sigma_8, n_s, h, \Omega_m, \Omega_b) =(0.82,0.97, 0.7,0.279, 0.046)$.

The shear and convergence maps come in the form of \textsc{Healpix}\footnote{\url{http://healpix.sf.net}} \citep{healpix,healpy} maps with resolution \texttt{NSIDE} = 4096. We first produced shear maps for each of the tomographic bins by averaging the shear snapshots weighted by the redshift distributions of the bins. To this aim, we used the approximate DES Y3 redshift distributions (\citet*{Myles_Alarcon_2021}). Galaxy catalogs are then created by sampling the simulated shear maps at the positions of real DES Y3 galaxies, matching their number density. While, in principle, shape noise can be added to the mock in order to closely match the real data specifications, we do not include it in our mock and instead regard simulation measurements as simple theory estimates.    

\section{Three-point Shear Correlations}\label{sec: theory}

We now describe the basic theory of the higher order correlations we are interested in, the estimator methods that are applied to the simulated and observed data described in the previous section, as well as data covariance matrix estimates based on jackknife.

\subsection{Theory Basics}\label{sec: theory basics}

Second order statistics  (two-point correlation functions, power spectra, second moments etc.) contain only the Gaussian part of the shear field. To probe non-Gaussian information, one has to appeal to higher-order statistics. 
We focus here on lensing 3rd order correlations. A fundamental aspect of these correlations is that they are projections of the matter bispectrum under some lensing kernel, so we take that as our starting point.

We first define the matter bispectrum $B_\delta(\boldsymbol{k}_1,\boldsymbol{k}_2,\boldsymbol{k}_3)$, that is, the Fourier transform of 3-point correlations of matter overdensities $\delta(\boldsymbol{k})$ in wavenumbers $\boldsymbol{k}$:
\begin{equation}\label{eq: matter bispectrum}
    \left\langle \delta(\boldsymbol{k}_{1})\delta(\boldsymbol{k}_{2})\delta(\boldsymbol{k}_{3})\right\rangle =B_{\delta}(k_{1},k_{2},k_{3})\delta_{\textrm{D}}(\boldsymbol{k}_{1}+\boldsymbol{k}_{2}+\boldsymbol{k}_{3}),
\end{equation}
where the Dirac delta $\delta_\textrm{D}$ enforces the bispectrum definition over wavenumbers $\boldsymbol{k}_i$ forming triangles, 
though with statistical isotropy the dependence is only on the magnitude of the modes $k_1$, $k_2$ and $k_3$ of the triangle.  The matter fluctuations give rise to a lensing signal that depends on the redshift distribution of the sources along a unit line-of-sight $\boldsymbol{\hat{n}}$. This is quantified in real space by the lensing convergence $\kappa(\boldsymbol{\hat{n}})$:
\begin{equation}
\kappa(\boldsymbol{\hat{n}})=\int_{0}^{\infty}dz\,W(\chi)\delta(\boldsymbol{\hat{n}},\chi),
\end{equation}
where $\chi=\chi(z)$ is the comoving distance to redshift $z$ and the lensing efficiency along the line-of-sight is 
\begin{equation}\label{eq: lensing kernel W}
W(\chi)=\frac{3\Omega_{\mathrm{m}}H_{0}^{2}}{2c^{2}}\frac{\chi}{a(\chi)}\int_{\chi}^{\infty}d\chi'\,n\left(z(\chi')\right)\frac{dz}{d\chi'}\frac{\chi'-\chi}{\chi'},
\end{equation}
where $\Omega_\textrm{m}$ is the matter density at redshift $z=0$, $H_0=100h$ km/s/Mpc is the Hubble parameter, $a$ is the scale factor, $n(z)$ is the normalized redshift distribution of sources, and $c$ is the speed of light. Under this lensing kernel, the 3-dimensional matter bispectrum in eq. (\ref{eq: matter bispectrum}) can be projected down to the 2-dimensional harmonic space convergence bispectrum using the Limber approximation \citep{Limber53,Limber_LoVerde2008}:
\begin{equation}\label{eq: convergence bispectrum}
    B_{\kappa}\left(\boldsymbol{\ell}_{1},\boldsymbol{\ell}_{2},\boldsymbol{\ell}_{3}\right)=\int_{0}^{\infty}d\chi\,\frac{W(\chi)^{3}}{\chi^{4}}B_{\delta}\left(\boldsymbol{k}_{1},\boldsymbol{k}_{2},\boldsymbol{k}_{3};\chi\right)
\end{equation}

With a weak lensing survey, we can probe the shear field at the positions of source galaxies and quantify its statistics with the lensing bispectrum above. We can define the spin-2 shear field along some direction (e.g., a line connecting two source galaxies) as $\gamma(\boldsymbol{\theta})=\gamma_t(\boldsymbol{\theta})+i\gamma_\times(\boldsymbol{\theta})$, where $\gamma_t$ is the shear component oriented perpendicularly with respect to that direction, $\gamma_\times$ is the 45$^{o}$ orientation, and $\boldsymbol{\theta}$ are vectors on the plane of the sky with magnitude $\theta$. A natural choice for two-point correlations of the shear field is to take the direction $\boldsymbol{\theta}$ to be that of the line separating a pair of source galaxies, in which case these correlations are given by 
\begin{equation}\label{eq: xipm definition}
    \xi_\pm(\theta)=\left\langle \gamma_{t}\gamma_{t}\right\rangle(\theta) \pm\left\langle \gamma_{\times}\gamma_{\times}\right\rangle(\theta) \equiv \gamma_{tt}\pm\gamma_{\times\times},
\end{equation}
with the angle brackets denoting averages taken over all possible pairs of galaxies, and where the right-hand equivalence introduces a shorthand notation for the multiplication of shears. 

While the choice for an orientation of shear projections in the three-point case is less obvious (e.g., the orthocenter of the triangle, or the side directions, etc.), there are ``natural components'' of cosmic shear with rotation and invariance properties analogous to $\xi_\pm$ that we can utilize \citep{Schneider2003_natural_components} [hereafter SL03]. We follow SL03 and define:

\begin{align}\label{eq: Gammas in terms of projections}
\nonumber\Gamma_{0}\equiv\left\langle \gamma(\boldsymbol{\theta}_{1})\gamma(\boldsymbol{\theta}_{2})\gamma(\boldsymbol{\theta}_{3})\right\rangle  & =\gamma_{ttt}-\gamma_{t\times\times}-\gamma_{\times t\times}-\gamma_{\times\times t}\\
 & +i\left[\gamma_{tt\times}+\gamma_{t\times t}+\gamma_{\times tt}-\gamma_{\times\times\times}\right],\\
\nonumber\Gamma_{1}\equiv\left\langle \gamma^{*}(\boldsymbol{\theta}_{1})\gamma(\boldsymbol{\theta}_{2})\gamma(\boldsymbol{\theta}_{3})\right\rangle  & =\gamma_{ttt}-\gamma_{t\times\times}+\gamma_{\times t\times}+\gamma_{\times\times t}\\
& +i\left[\gamma_{tt\times}+\gamma_{t\times t}-\gamma_{\times tt}+\gamma_{\times\times\times}\right],\\
\nonumber\Gamma_{2}\equiv\left\langle \gamma(\boldsymbol{\theta}_{1})\gamma^{*}(\boldsymbol{\theta}_{2})\gamma(\boldsymbol{\theta}_{3})\right\rangle  & =\gamma_{ttt}+\gamma_{t\times\times}-\gamma_{\times t\times}+\gamma_{\times\times t}\\
& +i\left[\gamma_{tt\times}-\gamma_{t\times t}+\gamma_{\times tt}+\gamma_{\times\times\times}\right],\\
\nonumber\Gamma_{3}\equiv\left\langle \gamma(\boldsymbol{\theta}_{1})\gamma(\boldsymbol{\theta}_{2})\gamma^{*}(\boldsymbol{\theta}_{3})\right\rangle  & =\gamma_{ttt}+\gamma_{t\times\times}+\gamma_{\times t\times}-\gamma_{\times\times t}\\& +i\left[-\gamma_{tt\times}+\gamma_{t\times t}+\gamma_{\times tt}+\gamma_{\times\times\times}\right]\label{eq: Gamma_end}.
\end{align}

It has been shown by SL03 as well as by \cite{Schneider2002, Takada_Jain_2003_correlations} that, for general triangle configurations, all of the correlations above can be non-zero and their imaginary parts do not necessarily vanish. Parity invariance, however, implies that the $\Gamma_i$ for equilateral configurations are purely real (all terms with an odd number of $\times$-components vanish) and that some, but not all, imaginary components of these statistics for isosceles configurations vanish. The correlations above thus have a complex configuration dependence and can be divided into a total of 8 data vectors (the real and imaginary part of each $\Gamma_i$), and should contain the entire 3pt information in the shear field.  

The $\Gamma_i$ are connected to the convergence bispectrum in eq.~(\ref{eq: convergence bispectrum}) since, in harmonic space, the shear components can be written in terms of the convergence as $\gamma(\boldsymbol{\ell})=e^{2i\beta}\kappa(\boldsymbol{\ell})$, where $\beta$ is the polar angle of $\boldsymbol{\ell}$. The exact expressions for each $\Gamma_i$ in terms of the convergence bispectrum is worked out in detail in \cite{Schneider_etal_2005}; for brevity, we simply quote their result for $\Gamma_0$ in simplified notation:

\begin{align}\label{eq: Gamma0 in terms of bispectrum}
\Gamma_{0}\left(\theta_{1},\theta_{2},\theta_{3}\right) & =(2\pi)\int_{0}^{\infty}\frac{\ell_{1}d\ell_{1}}{(2\pi)^{2}}\int_{0}^{\infty}\frac{\ell_{2}d\ell_{2}}{(2\pi)^{2}}\nonumber\\
 & \times\int_{0}^{2\pi}d\phi\,B_{\kappa}\left(\ell_{1},\ell_{2},\phi\right)\sum_{j=1}^{3}e^{i\alpha_{j}}J_{6}(A_{j}),
\end{align}
where $J_6$ is the 6-th order Bessel function of the first kind, and $B_\kappa = B_\kappa(\ell_1,\ell_2,\phi)$ due to statistical isotropy, with $\phi$ ithe polar angle between $\boldsymbol{\ell}_1$ and $\boldsymbol{\ell}_2$. We refer readers to \cite{Schneider_etal_2005} for the definitions of the coefficients $\alpha_j$ and $A_j=A_j\left(\theta_{1},\theta_{2},\theta_{3}\right)$ (see their eq. 15).

The shear field can also be decomposed into a different pair of statistics: the mass aperture statistic $M_\textrm{ap}$ and its cross-component $M_\times$ \citep{Schneider_etal_1998,Crittenden_etal_2002}. The mass aperture term is generally defined as a filtered version of the convergence $\kappa$:
\begin{equation}
    M_{\textrm{ap}}(\theta)=\int d^{2}\boldsymbol{r}U_{\theta}(r)\kappa(\boldsymbol{r}),
\end{equation}
and we can also introduce it in terms of the tangential  shear in circular apertures plus a cross-component shear term (expected to be null for an $E$-mode field) as:
\begin{align}\label{eq: M definition}
M(\theta) & =M_{\textrm{ap}}(\theta)+iM_{\times}(\theta)\nonumber\\
 & =\int d^{2}\boldsymbol{r}Q_{\theta}(r)\gamma_{t}(\boldsymbol{r})+i\int d^{2}\boldsymbol{r}Q_{\theta}(r)\gamma_{\times}(\boldsymbol{r}),
\end{align}
where again $\theta$ is the magnitude of a planar vector (an ``aperture radius'' over which the integrals above are computed), and $\boldsymbol{r}$ is a vector on the plane of the sky. 

There is some freedom in defining the filter functions $U_\theta (r)$ and $Q_\theta (r)$, but in this work we stick to the form proposed by \cite{Crittenden_etal_2002}:
\begin{equation}\label{eq: crittenden filter}
U_{\theta}(r)=\frac{1}{2\pi\theta^{2}}\left(1-\frac{r^{2}}{2\theta^{2}}\right)\exp\left(-\frac{r^{2}}{2\theta^{2}}\right),
\end{equation}
\begin{align}
Q_{\theta}(r) & =-U_{\theta}(r)+\frac{2}{r^{2}}\int_{0}^{r}r'\,dr'\,U_{\theta}(r')\\
 & =\frac{r^{2}}{4\pi\theta^{4}}\exp\left(-\frac{r^{2}}{2\theta^{2}}\right),\label{eq: Udef}
\end{align}
for an aperture of radius $\theta$. The statistics defined by eqs. (\ref{eq: M definition})-(\ref{eq: Udef}) have several interesting properties which have been explored in the literature \citep{Crittenden_etal_2002, Schneider_etal_2005, Kilbinger_Schneider_2005}. In particular, $M_\textrm{ap}$ and $M_\times$ cleanly separate, respectively, E- and B-modes of the shear field \citep{Shi14} and offer a relatively compact weighting over angular scales (note that the filter $Q_\theta(r)$ can be significantly non-zero for radii $r$ up to a factor of a few larger than the nominal aperture $\theta$, a feature we will come back to later). Additionally, these forms are mathematically tractable as they mainly involve Gaussian integrals. The ease of integration means that the connection between the third-order correlation of the mass aperture and the bispectrum is  straightforward. Again following \cite{Schneider_etal_2005}, we have:
\begin{align}\label{eq: Map3 in terms of bispectrum}
\left\langle M_{\textrm{ap}}^{3}\right\rangle (\theta_{1},\theta_{2},\theta_{3}) & =\frac{3}{(2\pi)^{3}}\int_{0}^{\infty}\ell_{1}d\ell_{1}\int_{0}^{\infty}\ell_{2}d\ell_{2}\int_{0}^{2\pi}d\phi\nonumber\\
 & \times B_\kappa(\ell_{1},\ell_{2},\phi)\tilde{U}\left(\theta_{1}\ell_{1}\right)\tilde{U}\left(\theta_{2}\ell_{2}\right)\tilde{U}\left(\theta_{3}\ell'\right),
\end{align}
where $\tilde{U}(x)=(x^{2}/2)e^{-x^{2}/2}$ is the Fourier transform of the filter $U_\theta(r)$ in eq. (\ref{eq: crittenden filter}) and $\ell'=\sqrt{\ell_{1}^{2}+\ell_{2}^{2}+2\ell_{1}\ell_{2}\cos\phi}$. The relatively compact weighting over $\ell$ multipoles provided by the filter and the absence of fast oscillatory functions in eq. (\ref{eq: Map3 in terms of bispectrum}) compared to eq. (\ref{eq: Gamma0 in terms of bispectrum}) make it a computationally tractable tool for theory predictions leading to cosmology, and indeed it has been a preferred statistic in the literature for cosmological constraints employing real space shear correlations \citep{JBJ04, Semboloni_etal_2011, Fu_etal_2014}. 

As a data vector, $\left\langle M_{\textrm{ap}}^{3}\right\rangle (\theta_{1},\theta_{2},\theta_{3})$ is easily tractable because it contains all three-point $E$-mode information in the field over all triangle configurations, as opposed to the complex splitting of the signal across the 8 non-zero $\Gamma_i(\theta_{1},\theta_{2},\theta_{3})$'s. We will also obtain measurements in the special case $\theta_1=\theta_2=\theta_3=\theta$ so that $\left\langle M_{\textrm{ap}}^{3}\right\rangle = \left\langle M_{\textrm{ap}}^{3}\right\rangle (\theta)$, which means all aperture radii are the same (though still accounting for different triangle configurations inside the apertures, not to be confused with a strict equilateral assumption). A schematic example of the angle variables used above and in Sec. \ref{sec: estimator} below is shown in Fig. \ref{fig: triangle notations} (Appendix \ref{appendix: coordinates}).  

It is interesting to consider, additionally, that as structure in the universe becomes more non-Gaussian at lower redshifts, the third order moments of the 3-dimensional density field should increase towards $z\to0$. For lensing fields, projection along the line of sight must also be included, and the evolution of non-Gaussian features is  quantified via the reduced \textit{skewness}  $S(\theta; z)$ \citep{Schneider_etal_1998}, showing the amplitude of the third moment relative to the second moment: 
\begin{equation}\label{eq: reduced map}
    S(\theta;z)=\frac{\left\langle M^3_\mathrm{ap}(z)\right\rangle}{\left\langle M^2_\mathrm{ap}(z)\right\rangle^2} (\theta),
\end{equation}
which is tightly related to the usual definition of the reduced bispectrum in terms of $B(k_1, k_2, k_3)/[P(k_1)P(k_2)+\mathrm{perm.}]$ \citep{Cooray_Sheth_2002}. This ratio encapsulates the contribution of non-Gaussian statistics to our low-redshift lensing data, arising predominantly from nonlinear structure formation at the scales considered in this work.

\subsection{Estimating $\Gamma_i$ and  $\left\langle M_\textrm{ap}^3 \right\rangle$}\label{sec: estimator}

Motivated by the connection between theory and observables in Sec. \ref{sec: theory basics} above, we now turn to the main objective of this work: to obtain and validate a measurement of shear correlations $\Gamma_i$ and $\left\langle M_{\textrm{ap}}^{3}\right\rangle$.

Our starting point is to measure the $\Gamma_i$'s. Their most straightforward data estimator is not conceptually different from estimating the usual 2pt statistics $\xi_\pm(\theta)$ in eq. (\ref{eq: xipm definition}). It relies on counting triplets (or pairs in the 2pt case) of galaxies in the survey, and accumulating the product of their shears in tangential and crossed orientations. So, for a catalog with ellipticities $\boldsymbol{e}=e_t+ie_\times$ with per-galaxy weights $w$, the estimator $\hat{\Gamma}_0$, for example,  is
\begin{equation}\label{eq: direct estimator}
    \hat{\Gamma}_{0}=\frac{\sum_{ijk}w_{i}w_{j}w_{k}\boldsymbol{e}_{i}\boldsymbol{e}_{j}\boldsymbol{e}_{k}}{\sum_{ijk}w_{i}w_{j}w_{k}},
\end{equation}
where the sum ($ijk$) runs over all galaxy triplets. In DES Y3, the weighting $w$ is given by the inverse variance of the ellipticity estimates in \textsc{Metacalibration} (see \citet*{Gatti2021} Sec. 4.3), and the ellipticities $\boldsymbol{e}$ are mean-subtracted and divided by the combination of shear and selection responses\footnote{Example usage of the DES Y3 shear catalogs is provided in  \url{https://github.com/des-science/DESY3Cats/}}.
Similar to the two-point $\xi_\pm$ case, this estimator is largely unaffected by masking and geometry of the survey.

For the other statistic, $\left\langle M_{\textrm{ap}}^{3}\right\rangle$, there are at least two conceptually different estimators. One relies on sampling apertures over the survey footprint and averaging over the tangential and cross components, directly probing integrals on the right-hand side of equation (\ref{eq: M definition}) as proposed by \cite{Schneider_etal_1998}. One of the main benefits of this method is that the estimation runtime can be made very fast  \citep{Porth2020}, and consequently it becomes feasible to obtain empirical survey covariance matrices of nearly arbitrary order in the mass aperture moments \citep{Porth2021}. A potential drawback of this estimator, however, is that survey masks, holes, edges and other common observational issues in real data can potentially bias the mass aperture estimate. 

A second method, which is our favored choice for the present work and was originally proposed by \cite{Schneider2002} and \cite{Crittenden_etal_2002}, relies on estimating the $n$-point statistics of the aperture mass by integrating over the $n$-point shear correlations themselves, as estimated from data. It was shown by \cite{JBJ04} that, by assuming the filtering function of \cite{Crittenden_etal_2002}, one obtains concise expressions for the $M(\theta)$ integration:   
\begin{equation}\label{eq: M3 integral}
    \left\langle M^{3}\right\rangle (\theta)=\int\frac{s\,ds}{\theta^{2}}\int\frac{d^{2}\boldsymbol{t}'}{2\pi\theta^{2}}\Gamma_{0}(s,\boldsymbol{t}')T_{0}\left(\frac{s}{\theta},\frac{\boldsymbol{t}'}{\theta}\right),
\end{equation}
\begin{equation}\label{eq: M2M* integral}
    \left\langle M^{2}M^{*}\right\rangle (\theta)=\int\frac{s\,ds}{\theta^{2}}\int\frac{d^{2}\boldsymbol{t}'}{2\pi\theta^{2}}\Gamma_{1}(s,\boldsymbol{t}')T_{1}\left(\frac{s}{\theta},\frac{\boldsymbol{t}'}{\theta}\right),
\end{equation}
where we have used the special case $\theta=\theta_1=\theta_2=\theta_3$, where $s$ and $\boldsymbol{t}'$ are triangle sides as defined in eq.(\ref{eq: q s t definition}), 
and the functions $T_0$ and $T_1$ are defined in eqs.(\ref{eq: T0 definition}) and (\ref{eq: T1 definition}) (see Appendix \ref{appendix: coordinates}). The separate tangential and cross components $\left\langle M_\textrm{ap}^3\right\rangle$ and $\left\langle M_\times^3 \right\rangle$ can be written as linear combinations of the $\left\langle M^{3}\right\rangle$ and $\left\langle M^{2}M^{*}\right\rangle$ defined above. In particular, with $\mathcal{R}$ denoting the real part of an imaginary quantity, we have
\begin{equation}\label{eq: Map3 in terms of M}
    \left\langle M_{\textrm{ap}}^{3}\right\rangle (\theta)=\frac{1}{4}\mathcal{R}\left[3\left\langle M^{2}M^{*}\right\rangle (\theta)+\left\langle M^{3}\right\rangle (\theta)\right].
\end{equation}

We utilize \textsc{TreeCorr} \citep{JBJ04} in order to estimate the quantities in eqs. (\ref{eq: direct estimator})-(\ref{eq: Map3 in terms of M}) above. \textsc{TreeCorr} is an efficient tree-based algorithm for computing 2pt and 3pt correlation functions in real space data. The estimator follows closely equations (\ref{eq: direct estimator}), (\ref{eq: M3 integral}) and (\ref{eq: M2M* integral}) in the sense that galaxy shears are first aggregated by their triangle configuration and side lengths, and in a post-processing step the $\Gamma_i$ are integrated over with the $T_{0,1}$ functions to obtain $M_{\textrm{ap}/\times}$. The base algorithm itself is the same utilized for correlation function measurements in the two-point DES Y3 cosmology results \citep{y3-3x2ptkp}. We refer the reader to the source code and documentation webpage for more information\footnote{\texttt{https://github.com/rmjarvis/TreeCorr}}.

Even with a highly efficient tree algorithm, we find that runtime is a limiting factor when computing 3pt correlations of the spin-2 shear fields in our data (see Sec. \ref{sec: estimator uncertainties} further below). Therefore, for all measurements presented in this work, we divide the survey (and simulation) footprints into 100 patches of nearly equal number of galaxies. With DES Y3 data, each patch contains about $N=1$M galaxies. The main advantage of this approach is to significantly reduce the number of galaxies dealt with in each measurement and to better parallelize it. 

We define the patch centers and assign galaxies to them using the $k$-means implementation in \textsc{TreeCorr}, which yields patches of roughly similar area $\gtrsim$40 deg$^2$ (a characteristic length $\gtrsim6$ deg). This choice is sub-optimal, because measuring correlations in finite patches of an otherwise contiguous area necessarily neglects the signal contributions coming from triangles formed by galaxies that lie in different patches. However, since the area of the DES Y3 footprint is large compared to the relatively small angular scales over which we present our measurements in Sec. \ref{sec: results}, this is not a significant issue. We return to this and other estimator tests in Sec. \ref{sec: estimator uncertainties}.        

Due to the angular binning performed by \textsc{TreeCorr}, for triangles of side lengths $d_3\leq d_2 \leq d_1$, we define, more conveniently
\begin{equation}\label{eq: medium}
    \theta_\mathrm{medium}=d_2
\end{equation} 
as a proxy to index the $\Gamma_i$ data vector, and unless explicitly noted otherwise we average over all triangles that fall within a bin around $\theta_\mathrm{medium}$. \textsc{TreeCorr} uses internal variables $u$ and $v$ (defined in eqs. \ref{eq: u definition} and \ref{eq: v definition}) that characterize triangles by their configuration (eg. squeezed or equilateral). We then estimate the mean 3pt signals for each natural component $i$ of $\Gamma_i(\theta_\mathrm{medium})$ 
via the weighted sample mean over the patches $\alpha$, with $\alpha\in[1,100]$: 
\begin{equation}\label{eq: mean weighted Gamma signal}
\Gamma_{i}(\theta_{\mathrm{medium}})=\frac{\sum_{\alpha}\sum_{uv}(1/\mathrm{Var}\left[\Gamma_{i,\alpha}\right])\Gamma_{i,\alpha}(\theta_{\mathrm{medium}},u,v)}{\sum_{\alpha}\sum_{uv}(1/\mathrm{Var}\left[\Gamma_{i,\alpha}\right])}
,
\end{equation}
where inverse-variance weights are estimated in the shape noise regime (more details in Sec. \ref{sec: JK cov}). Analogously, we compute the skewness of the mass aperture in each patch using eq. (\ref{eq: Map3 in terms of M}) and then combine them so the mean signal is 
\begin{equation}\label{eq: mean weighted Map3 signal}
\left\langle M_{\mathrm{ap}}^{3}\right\rangle (\theta_{1},\theta_{2},\theta_{3})=\frac{\sum_{\alpha}(1/\mathrm{Var}\left[M_{\mathrm{ap}}^{3}\right])\left\langle M_{\mathrm{ap}}^{3}\right\rangle _{\alpha}}{\sum_{\alpha}(1/\mathrm{Var}\left[M_{\mathrm{ap}}^{3}\right])}.
\end{equation}

\subsection{Covariance Matrix}\label{sec: JK cov}

With the computation of the measurement over $N=100$ patches of the DES Y3 data, we can readily obtain a jackknife estimate of the covariance matrix:
\begin{equation}
\mathrm{Cov}\left[\zeta(\theta_{i}),\zeta(\theta_{j})\right]=\frac{N-1}{N}\sum_{\alpha}\Delta \zeta_{\alpha}(\theta_{i}) \Delta \zeta_{\alpha}(\theta_{j})^T
\end{equation}
where $\zeta$ is the data vector of the statistic under consideration (\map$(\theta)$ or $\Gamma_i(\theta_\mathrm{medium})$ for instance), $\left\langle \zeta\right\rangle$ is its average value over the $N$ patches, and  $\Delta \zeta_{\alpha}\equiv \zeta_{\alpha}-\left\langle \zeta\right\rangle $. When inverting the covariance matrix, we also apply a ``Hartlap correction'' factor \citep{Hartlap, DodelsonSchneider2013, Sellentin2016, TaylorJoaKit2013} given by $(P-N-1)/(N-2)$ where $P$ is the dimension of the data vector and N is the number of patches ($P=7$ and 55 for \map$(\theta)$ and $\Gamma_i(\theta_\mathrm{medium})$ respectively, and $N=100$ in both cases). 

\begin{figure}
	\includegraphics[width=\columnwidth]{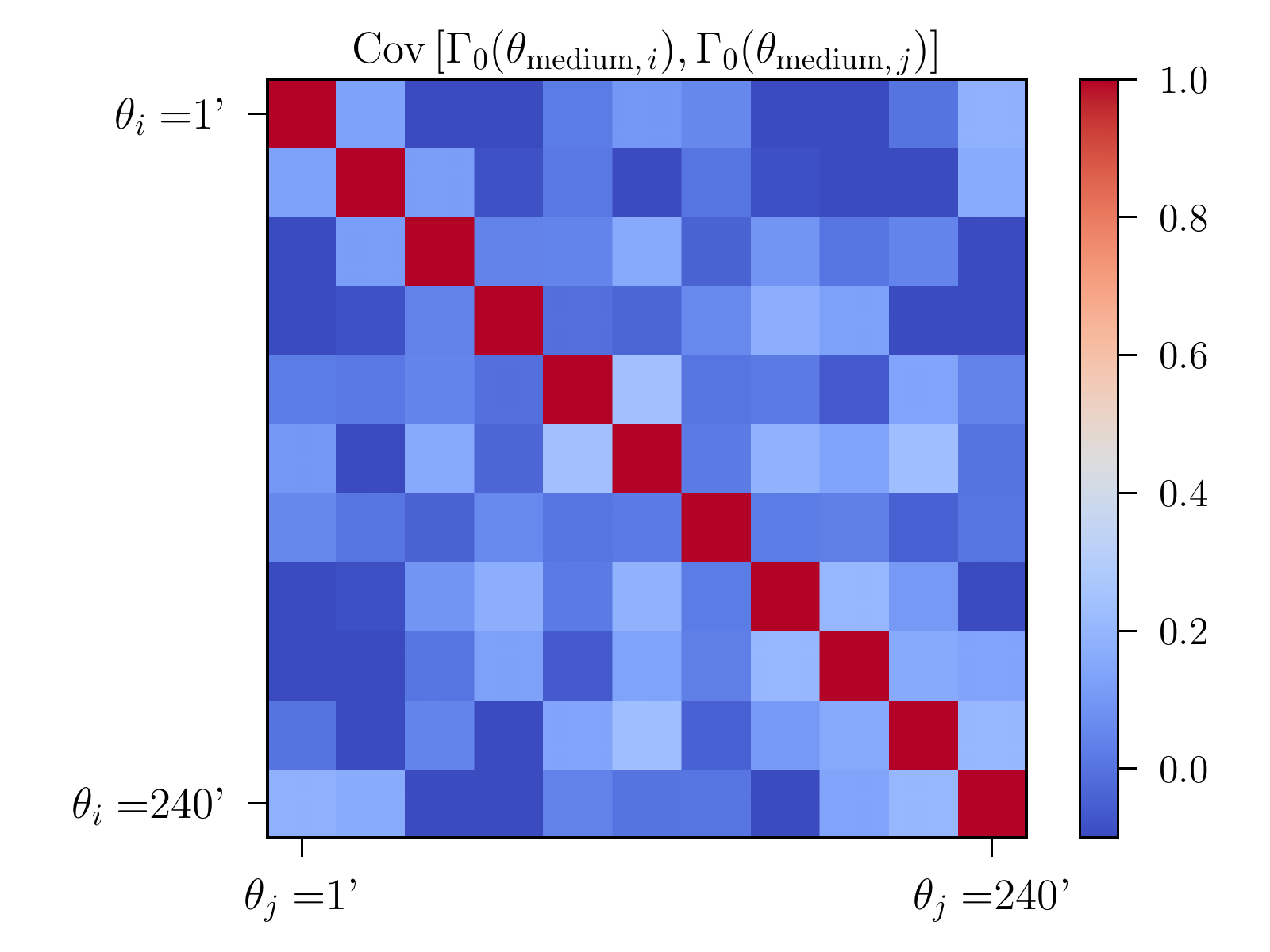}
	\includegraphics[width=\columnwidth]{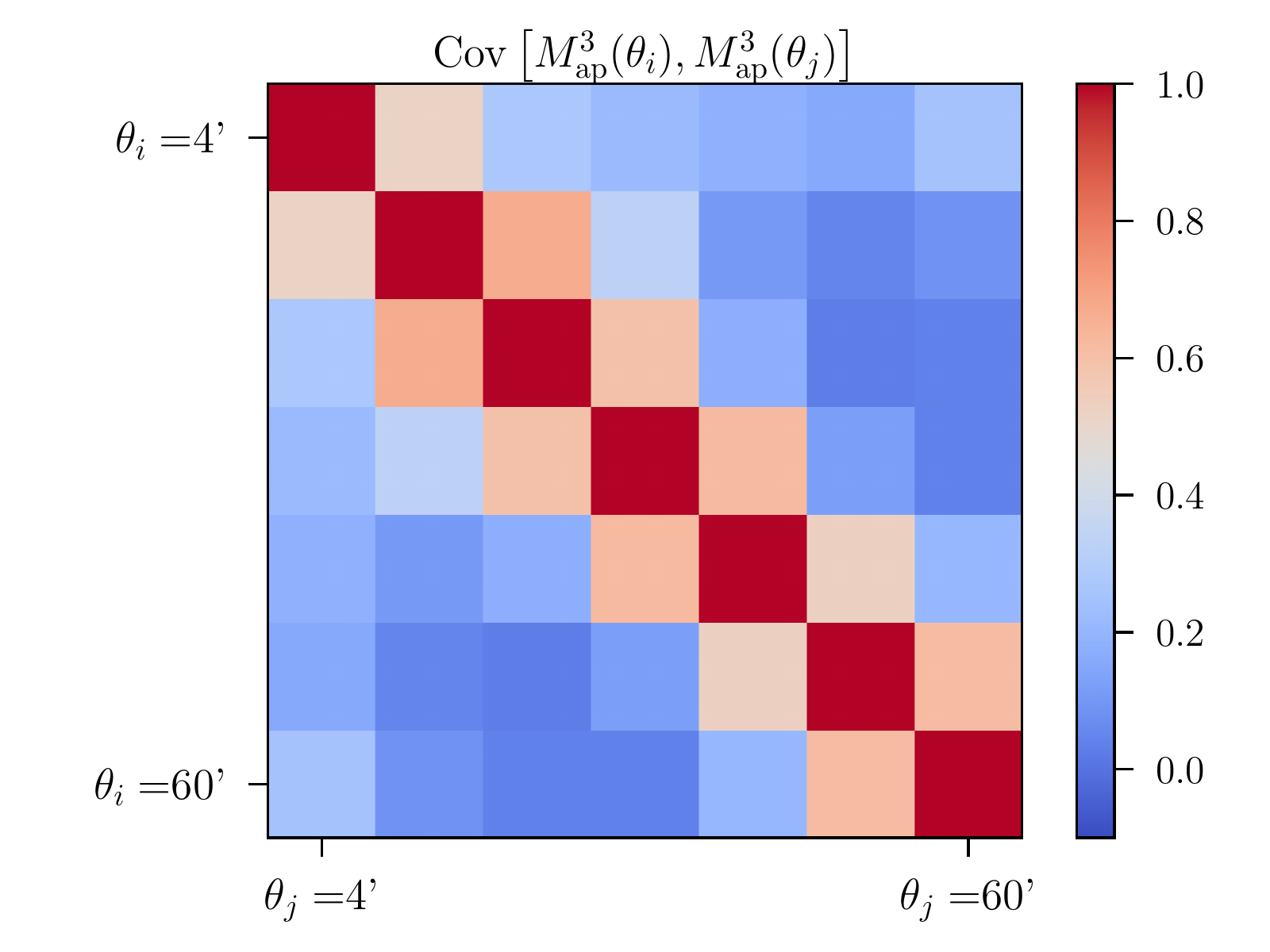}
	\includegraphics[width=\columnwidth]{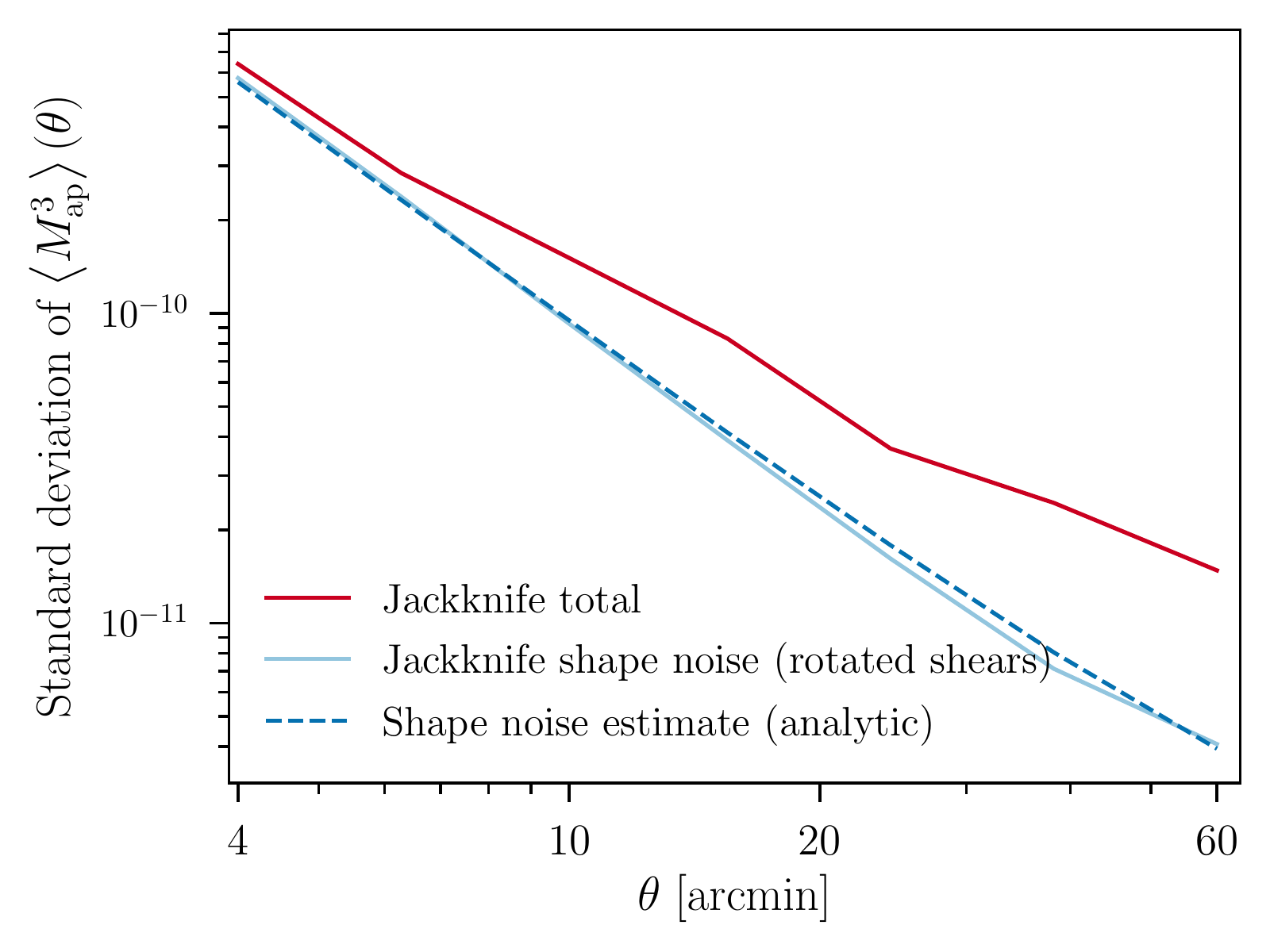}
    \caption{Normalized covariance matrix estimates for $\map$ (top panel), $\Gamma_0$ (middle panel) and $\map$ standard deviation diagonal (bottom panel). With the exception of the analytic shape noise estimate (dashed line in bottom panel), all other estimates are obtained from the jackknife measurements on DES Y3 data. We note that in most scales below around 20 arcmin, the shape noise contribution to the error bars is of around 50\% or more, and at larger scales the errors are dominated by cosmic variance.
    \label{fig: covariances}
    }
\end{figure} 

In Fig. \ref{fig: covariances}, we show the normalized covariance matrices (correlation matrices) for the \map$(\theta)$ and $\Gamma_0(\theta_{\mathrm{medium}})$ estimates which we present in the following Section. We additionally show, on the bottom panel of that Figure, how the diagonal \map~standard deviation compares with empirical and analytic estimates of the error in the shape noise dominated regime. We obtain an empirical estimate of the shape noise signal (light blue curve in Fig. \ref{fig: covariances}) by repeating the \map~measurement over patches in which each individual galaxy shear has been randomly rotated. This effectively cancels out the cosmic signal and variance, leaving us with an estimate of the shape noise that preserves any masking or geometry effects of the real data. We additionally overplot in that same panel an analytic estimate of shape noise. The analytic estimate comes from the propagation of the weighted variance of $\Gamma_i$ into \map, which in turn can be written as
\begin{equation}
    \mathrm{Var}\left[\mathcal{R}\left\{ \Gamma\right\} \right]=4\sigma_{e}^{6}\frac{\sum_{ijk}w_{i}^{2}w_{j}^{2}w_{k}^{2}}{\left(\sum_{ijk}w_{i}w_{j}w_{k}\right)^{2}}
\end{equation}
where $w$ are weights associated to the data ellipticities, $\sigma^2_e=\left\langle (e_i - \left\langle e_i \right\rangle)^2 \right\rangle$ is the variance of single-component ellipticities, and the sums $(ijk)$ run over all possible triplets of galaxies. We note that this reduces to $\mathrm{Var}\left[\mathcal{R}\left\{ \Gamma\right\} \right] = 4\sigma^6_e/N_\vartriangle$ for equal galaxy weighting, where  $N_\vartriangle$ is the number of triangles in a given angular bin. We find that, for values of $\theta$ less than $\sim$20 arcmin, the shape noise contributes $>50\%$ of the estimated error bars in \map. 

While jackknife covariances are known to be biased on scales that approach the characteristic length of an individual patch, the covariances we utilize should be reliable for the simple $S/N$ estimates at the relatively smaller angular scales studied in this work. Survey data covariances are generally difficult to obtain and can directly impact likelihood analyses, especially at the 3pt level \citep{Sato2013, Joachimi2009}. We therefore intend to further study the suitability of our existing jackknife matrices in a follow-up work focusing on the inference of cosmology constraints.

\section{Measurement Results}\label{sec: results}

\begin{figure*}
	\includegraphics[width=\columnwidth]{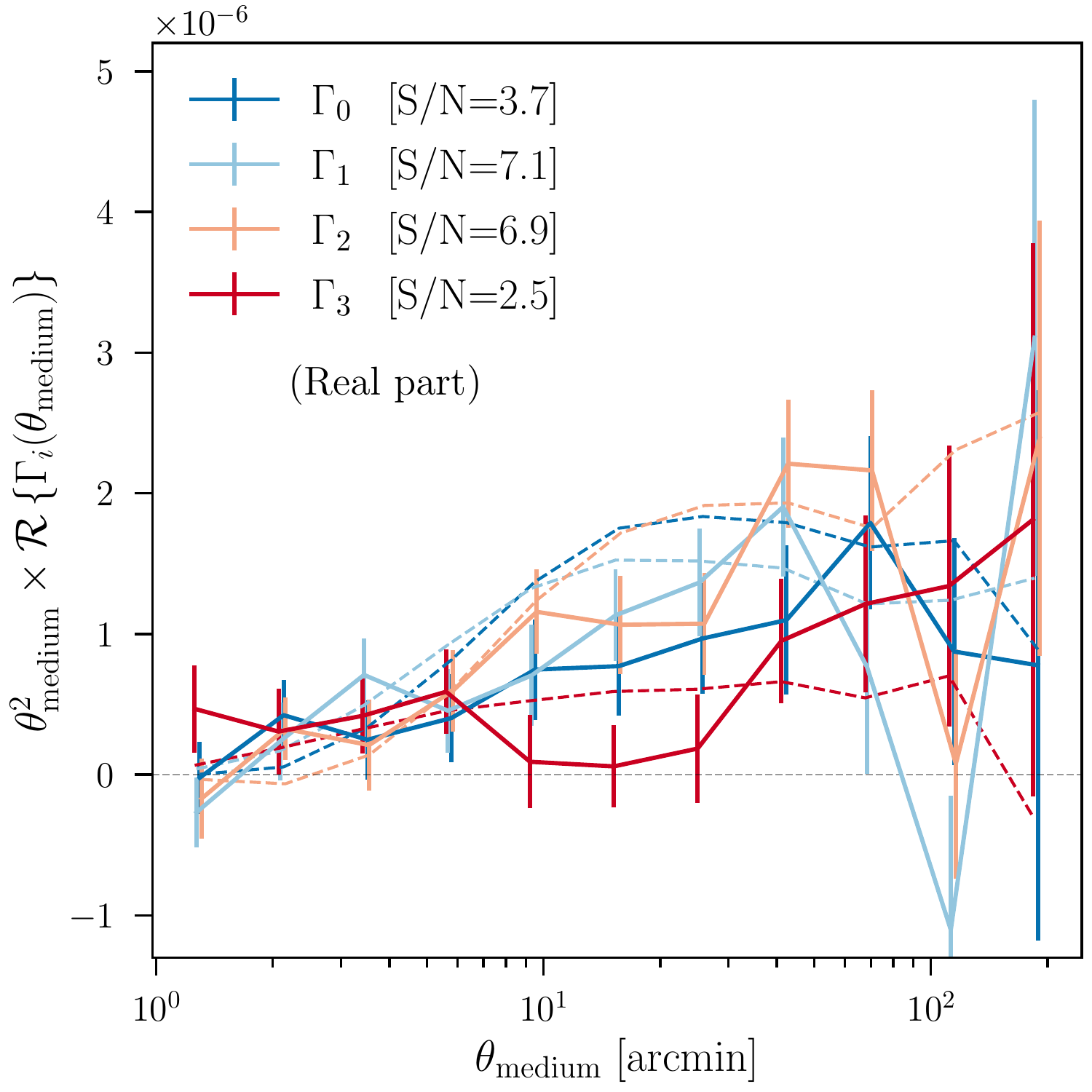}
	\includegraphics[width=\columnwidth]{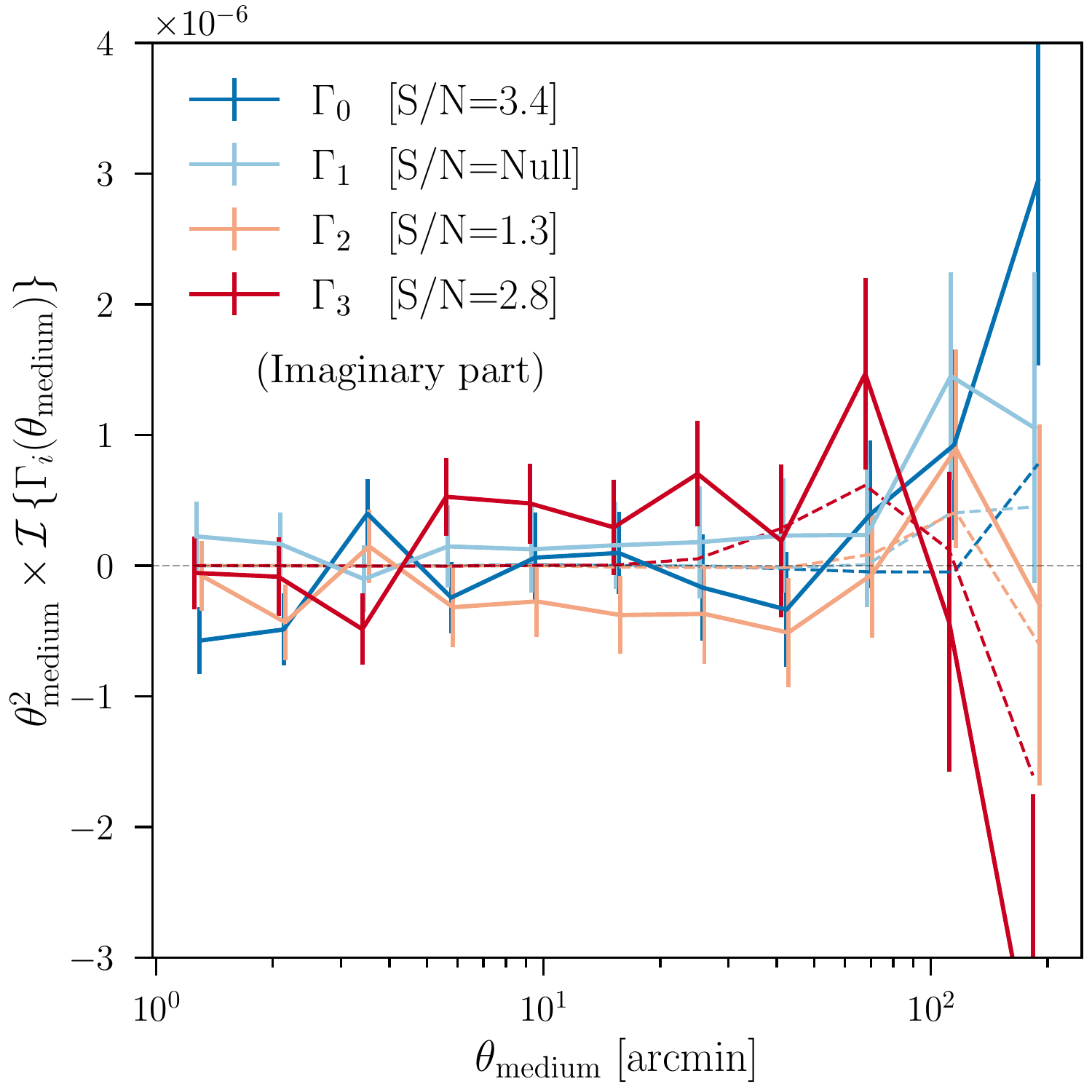}
    \caption{The non-tomographic 3rd order natural shear correlations $\Gamma_i$ in DES Y3 as a function of  angular scale in arcminutes of the medium-length side of triangles, $\theta_\mathrm{medium}$ (eq. \ref{eq: medium}). Solid lines correspond to averaged measurements over 100 patches of the DES Y3 footprint; error bars are estimated with a jackknife method. To guide the eye, the dashed lines show corresponding  measurements on a T17 N-body mock catalog (Sec. \ref{sec: takahashi}) with DES Y3-like redshift distributions but without shape noise and for an older set of cosmological parameters. Signal-to-noise (S/N) estimates are obtained with eq.(\ref{eq: SN definition}). \textbf{Left panel:} Real parts of the natural 3pt shear components. The lensing signal is distributed rather evenly across the 4 components, and for most of them the null-hypothesis is clearly rejected at high confidence. \textbf{Right panel:} Imaginary parts of the natural 3pt shear components, which are expected to be zero for certain triangle configurations (e.g., equilateral) but not in general, thus leading to smaller overall $S/N$.}
    \label{fig: non-tomographic}
\end{figure*}

We now apply the estimators defined in Sec. \ref{sec: estimator} to the DES Y3 data split into 100 patches. We measure the 3pt correlations $\Gamma_i$ within an angular range of $\theta_\textrm{medium}\in[1, 240]$ arcmin, approximately the same range of scales validated in DES Y3 for weak lensing applications. For $\Gamma_i$, angular bins in $\theta_\textrm{medium}$ are log-spaced (with 0.1 spacing, leading to 55 bins) and \textsc{TreeCorr}'s internal variables $u$ and $v$ are linearly-spaced (0.1 spacing, leading to respectively 10 and 20 bins; see Appendix \ref{appendix: coordinates}) to ensure stability of the integrals that lead to $M_\textrm{ap}$ moments. When plotting $\Gamma_i$ results and obtaining its covariance, we further average over every 5 bins in $\theta_\textrm{medium}$ for ease of visualization and to reduce noise. For the results on the $M_\textrm{ap}$ estimation, however, we focus on a narrower range of scales and limit aperture radii to the interval $\theta\in[4,60]$ arcmin in 7 bins, avoiding measurement biases that can arise if the aperture filtering in eq.(\ref{eq: Udef}) spans scales over which the $\Gamma_i$ were not obtained (further details in Sec. \ref{sec: estimator uncertainties}). 

We present the non-tomographic signal in Sec. \ref{sec: non-tomographic map3}, along with splits of triangles by configuration type, and then we divide our data into two tomographic bins in Sec. \ref{sec: tomographic map3}. In what follows, we define the signal-to-noise ($S/N$) of our detections as (see Appendix~\ref{snr_vector}, where this is derived)
\begin{equation}\label{eq: SN definition}
S/N\equiv\begin{cases}
\sqrt{\chi^{2}-N_{\textrm{d.o.f}}} & \textrm{if }\chi^{2}\geq N_{\textrm{d.o.f}}+1\\
\textrm{``Null''} & \textrm{otherwise}
\end{cases},
\end{equation}
where $N_{\mathrm{d.o.f}}$ is degrees of freedom (here the number of data points) and $\chi^{2}=\boldsymbol{d}^{T}C^{-1}\boldsymbol{d}$ with $\boldsymbol{d}$ representing the measurement vector and $C^{-1}$ representing the inverse data covariance. In the low signal-to-noise regime (which is the case for many of the null tests presented later), it may be that $\chi^2 < N_{\mathrm{d.o.f}}+1$, in which case  $S/N$ is less than $1.0$ or imaginary, which we consider a ``Null'' signal (consistent with no detection). Additionally, for practical purposes, we define a data vector to be \textit{significantly} rejecting the null-hypothesis (at $X\sigma$) if $S/N=X>2.5$, which as an equivalent $p$-value yields $p\lesssim0.01$.

\subsection{Non-tomographic 3pt Shear Signal}\label{sec: non-tomographic map3}

We first focus on the non-tomographic setting, treating all galaxies in the survey as if their line-of-sight distances belonged to a thin plane on the sky. We show the real and imaginary parts of the non-tomographic $\Gamma_i$ and their $S/N$ in Fig. \ref{fig: non-tomographic}. We report significant detections (ruling out the null-hypothesis at 2.5$\sigma$ or more) of the real parts of all natural shear components $\Gamma_i(\theta_\mathrm{medium})$, and an overall lower significance for their imaginary parts. This is expected since, in specific triangle configurations, but not generally all of them, the imaginary parts vanish due to parity conservation. We overplot measurements obtained from the T17 N-body mock with dashed lines as a guide to the eye, though it should not be expected that these curves serve as a fit to the data, which we return to below.

\begin{figure*}
	\includegraphics[width=\columnwidth]{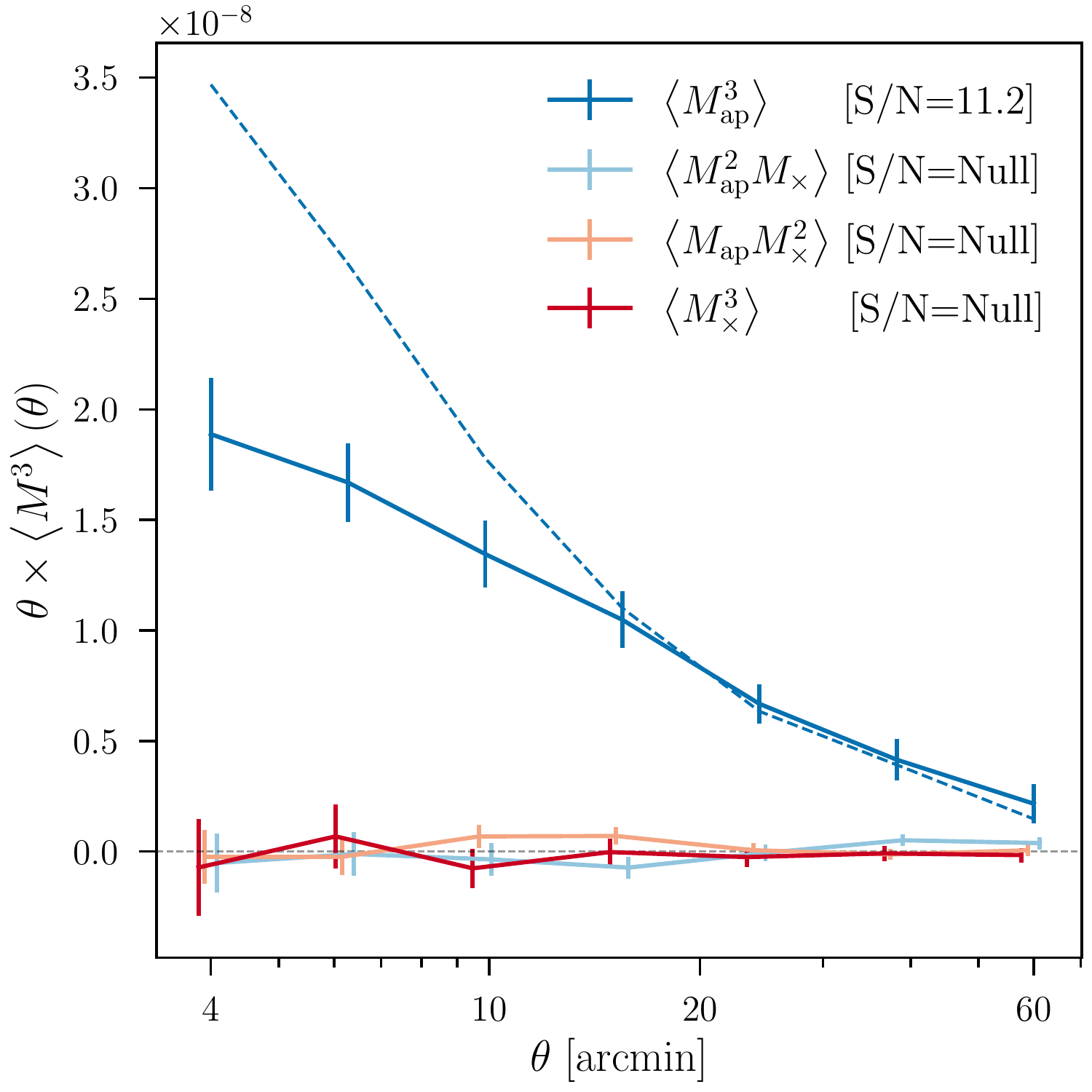}
	\includegraphics[width=\columnwidth]{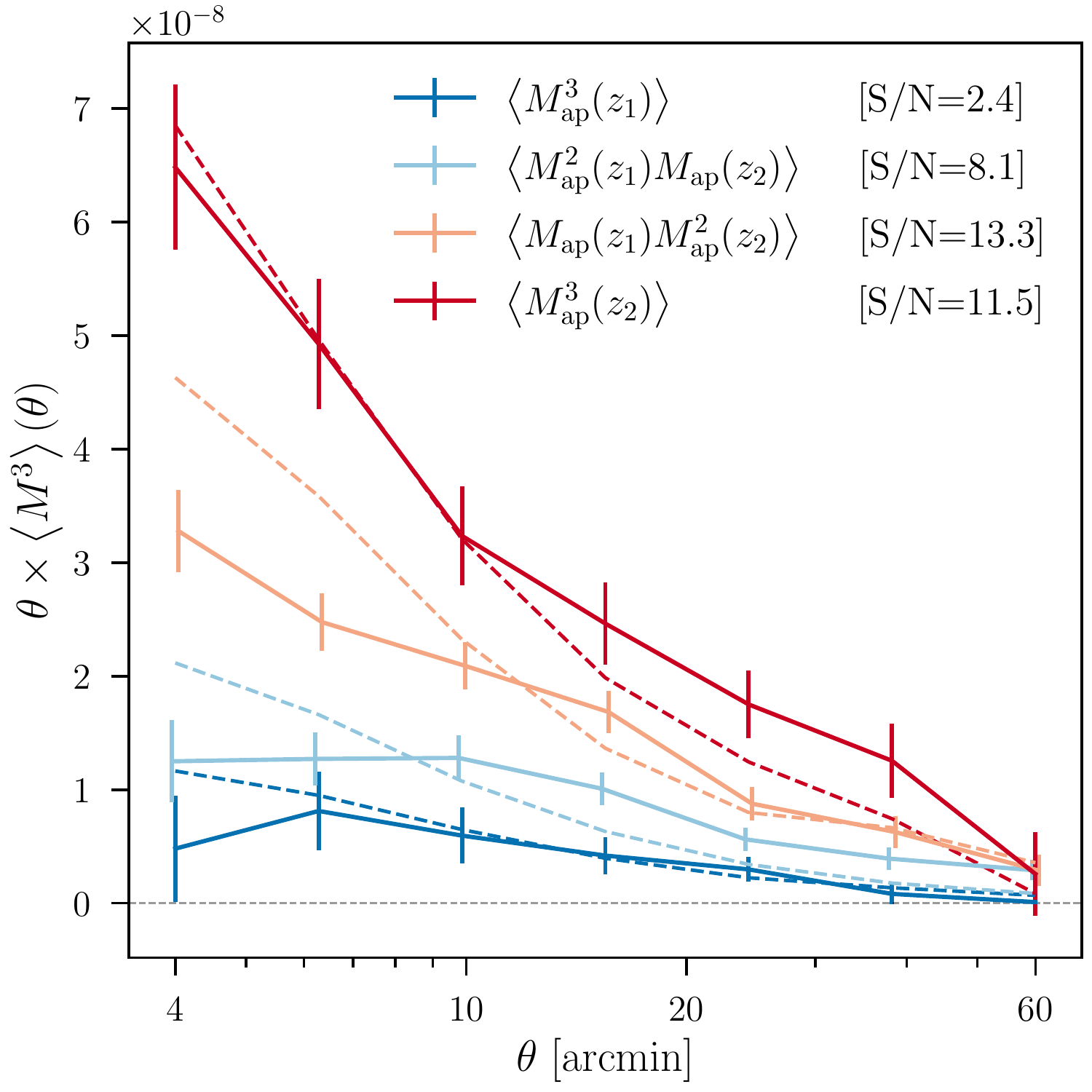}
    \caption{The 3rd order mass aperture correlations in DES Y3 for the special case of a single aperture radius $(\theta_1=\theta_2=\theta_3=\theta)$ in arcminutes.  Solid lines correspond to averaged measurements over 100 patches of the DES Y3 footprint, and error bars are estimated with a jackknife method. As in Fig. \ref{fig: non-tomographic},  dashed lines show measurements for the T17 N-body mock (Sec. \ref{sec: takahashi}), and S/N estimates are obtained with eq.(\ref{eq: SN definition}) . \textbf{Left panel:} non-tomographic $M_\mathrm{ap}$ and $M_\times$ moments. We find a significant detection of the pure $E$-mode term $\left\langle M^{3}_{\mathrm{ap}} \right\rangle $, and find the other combinations, which either violate parity or imply significant $B$-mode contamination, to be consistent with the null-hypothesis. \textbf{Right panel:} Tomographic mass aperture cross-correlations using DES Y3 data split into 2 wide redshift bins $z_1$ and $z_2$. We find a significant detection of the cross-correlations that include the higher redshift bin $z_2$, and the total combined data vector is detected with $15.0\sigma$ significance. }
    \label{fig: 1-aperture Map}
\end{figure*}

We additionally report a strong detection (ruling out the null hypothesis at more than 11$\sigma$) of the non-tomographic lensing $E$-mode term \map~in the left panel of Fig. \ref{fig: 1-aperture Map}, in the special case of a single aperture radius $\theta_1=\theta_2=\theta_3$. The higher $S/N$ of the mass aperture in comparison with individual $\Gamma_i$'s is in principle expected: the tangential projection of shears for a given triangle configuration contains a large fraction of the signal \citep{Takada_Jain_2002} and the \map~statistic sums over that projection across many configurations in an aperture $\theta$, while $\Gamma_i$ splits the contribution over a total of 8 independent correlations $\gamma_{abc}$ with $a,b,c\in[t,\times]$.

We find that the overall amplitude of the simulated and data  signals in both $\Gamma_i$ and \map~ closely resemble each other. A more careful assessment beyond the scope of this work would be necessary to verify whether discrepancies between solid and dashed lines in Fig \ref{fig: 1-aperture Map} imply our data are statistically rejecting the cosmology (or gravity-only implementation) of the T17 simulations. 

Several known factors could result in these differences: the difference in assumed cosmology, small scale astrophysical systematic effects, and shear calibration. Discerning between these factors would entail obtaining 3pt functions in the ensemble of 108 mocks in T17 as opposed to the single shape-noise free mock utilized in this work, a computationally expensive task (see Sec. \ref{sec: estimator uncertainties} for details on the estimator performance), and carrying out likelihood analyses over scales where the theory modeling is not excessively uncertain. We do note, however, that based on the left panel of Fig. \ref{fig: 1-aperture Map} the largest offsets are on small scales (roughly below $10'$) and result in a $\Delta\chi^2\approx40$ when comparing data and mock within $\left\langle M^3_{\mathrm{ap}}\right\rangle (\theta<10')$ for the non-tomographic case. Similarly, the tomographic measurements $\left\langle M_\mathrm{ap}(z_1)M_\mathrm{ap}(z_2)^2 \right\rangle (\theta<10')$ and $\left\langle M_\mathrm{ap}(z_1)^2 M_\mathrm{ap}(z_2) \right\rangle (\theta<10')$ over the same scales show a combined $\Delta\chi^2\approx 30$, so it may be possible that the origin of the non-tomographic discrepancy is driven by the redshift cross-correlations. This likely rules out strong baryonic feedback in the data as an explanation for the discrepancy (as that would also have shown up strongly in the lowest-redshift $\left\langle M_\mathrm{ap}(z_1)^3\right\rangle (\theta<10')$ for most feedback scenarios) as well as significant contributions from shear calibration bias (which would likely have appeared as a scale-independent offset affecting additionally the auto-redshift correlations). We leave further detailed explorations for a future work.

Comparing the $\left\langle M^3_{\mathrm{ap/}\times}\right\rangle (\theta)$ and $\Gamma_i(\theta_\mathrm{medium})$ statistics presented in Fig. \ref{fig: non-tomographic} and in the left panel of Fig. \ref{fig: 1-aperture Map} we find that they separate the signal contributions in different ways. While for general triangle configurations the $E$ and $B$ mode signals are split rather evenly between the $\Gamma_i$, they are more concentrated in $M_\mathrm{ap}$ as opposed to  $M_\times$. We will exploit this feature in more detail in Sec. \ref{sec: validation} as an assessment of systematics.

\begin{figure*}
	\includegraphics[width=\columnwidth]{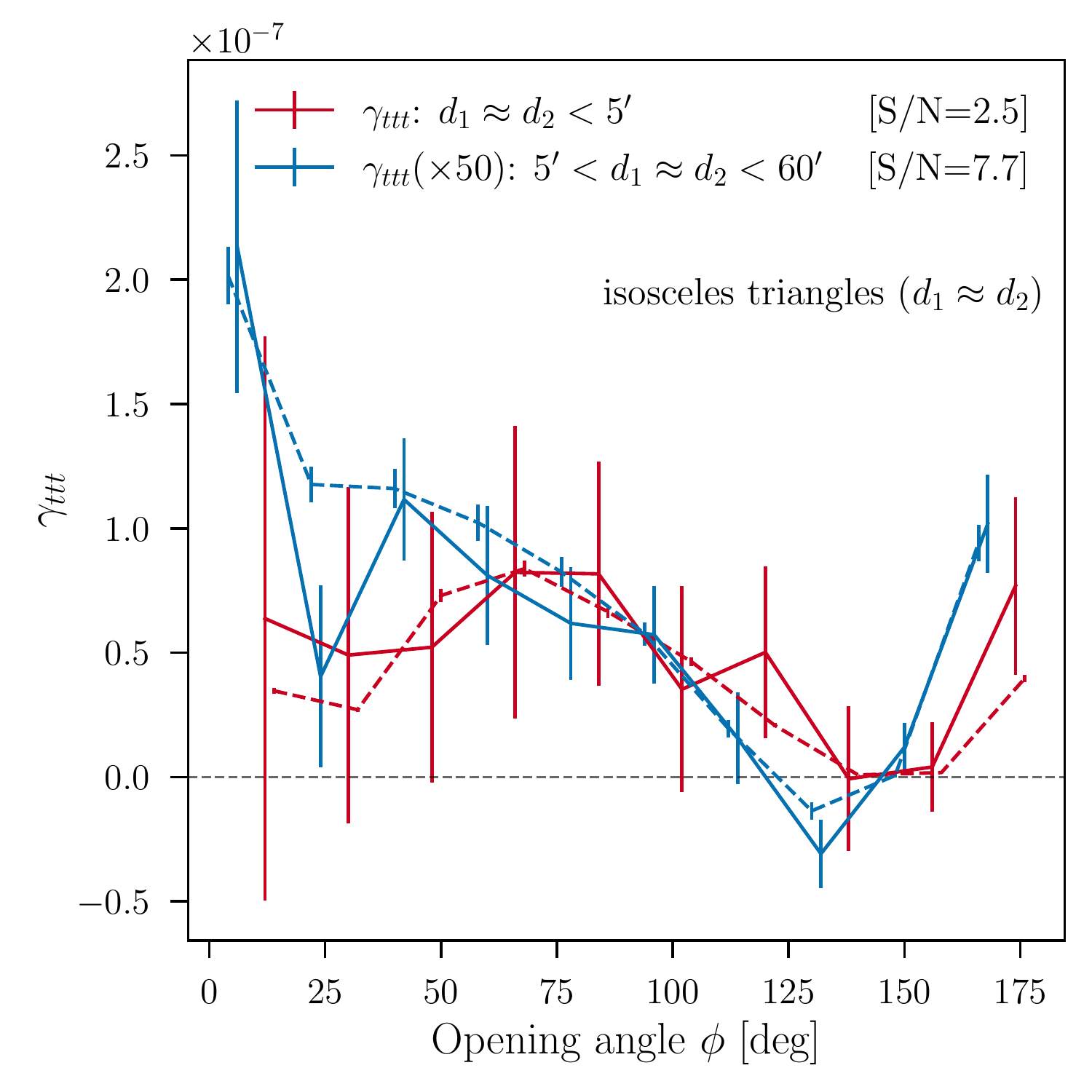}
	\includegraphics[width=\columnwidth]{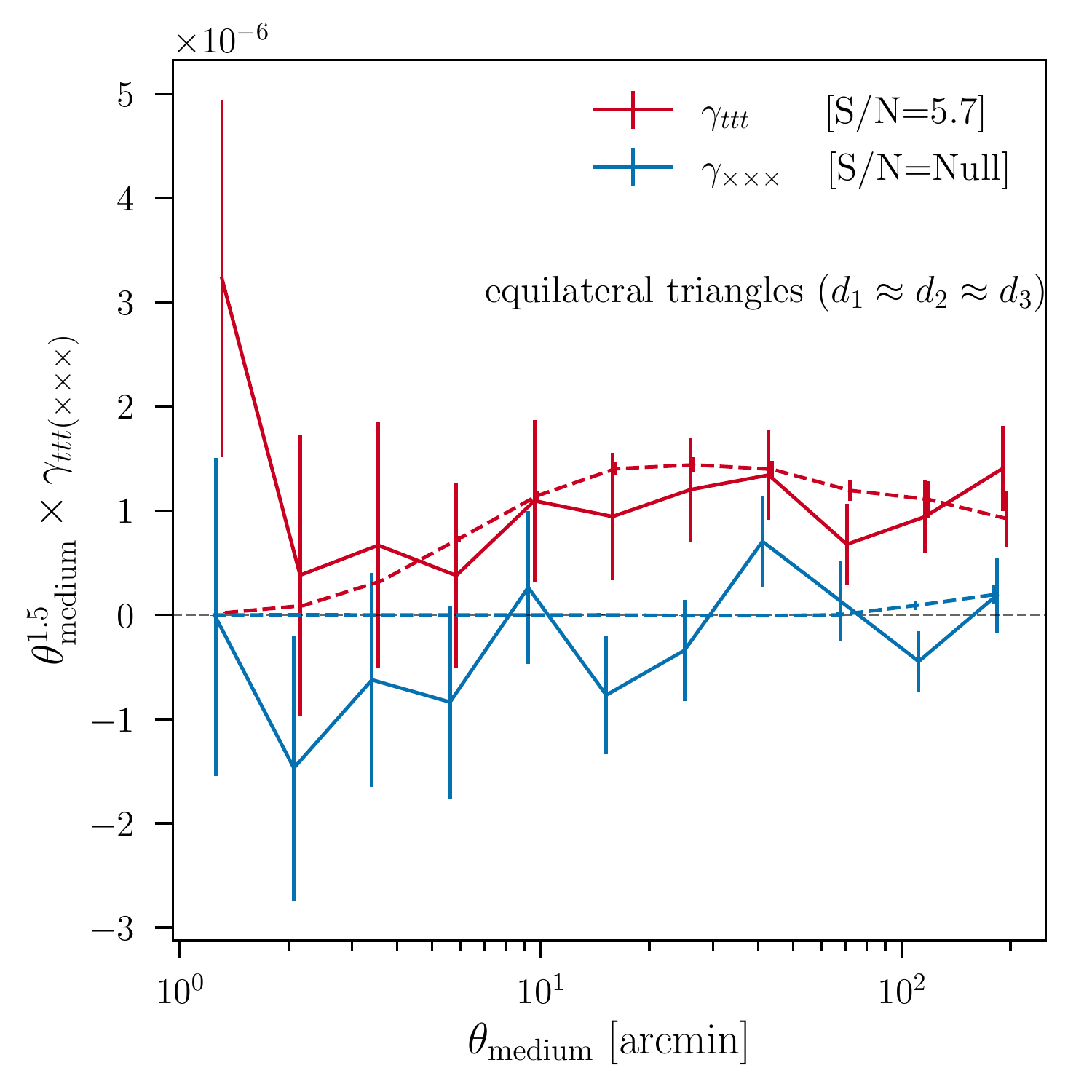}
    \caption{Three-point shear signals with different triangle configuration and scale dependences. Solid lines correspond to non-tomographic DES Y3 data measurements, and dashed lines are the same measurements made on a T17 N-body mock. \textbf{Left panel:} the purely tangential $\gamma_{ttt}$ component for isosceles triangles as a function of opening angle between sides $d_1$ and $d_2$. An oscillating pattern (see text) can be seen at both the small scales (red) as well as large-scale triangles (blue, multiplied by a factor of 50 for visualization). \textbf{Right panel:} purely tangential ($\gamma_{ttt}$) and cross-projections ($\gamma_{\times\times\times}$) of equilateral triangles. A detection is clear in the tangential case and, according to expectations the equilateral, odd-parity $\gamma_{\times\times\times}$ is consistent with zero.}
    \label{fig: configuration dependence}
\end{figure*}

\begin{figure}
	\includegraphics[width=1.0\columnwidth]{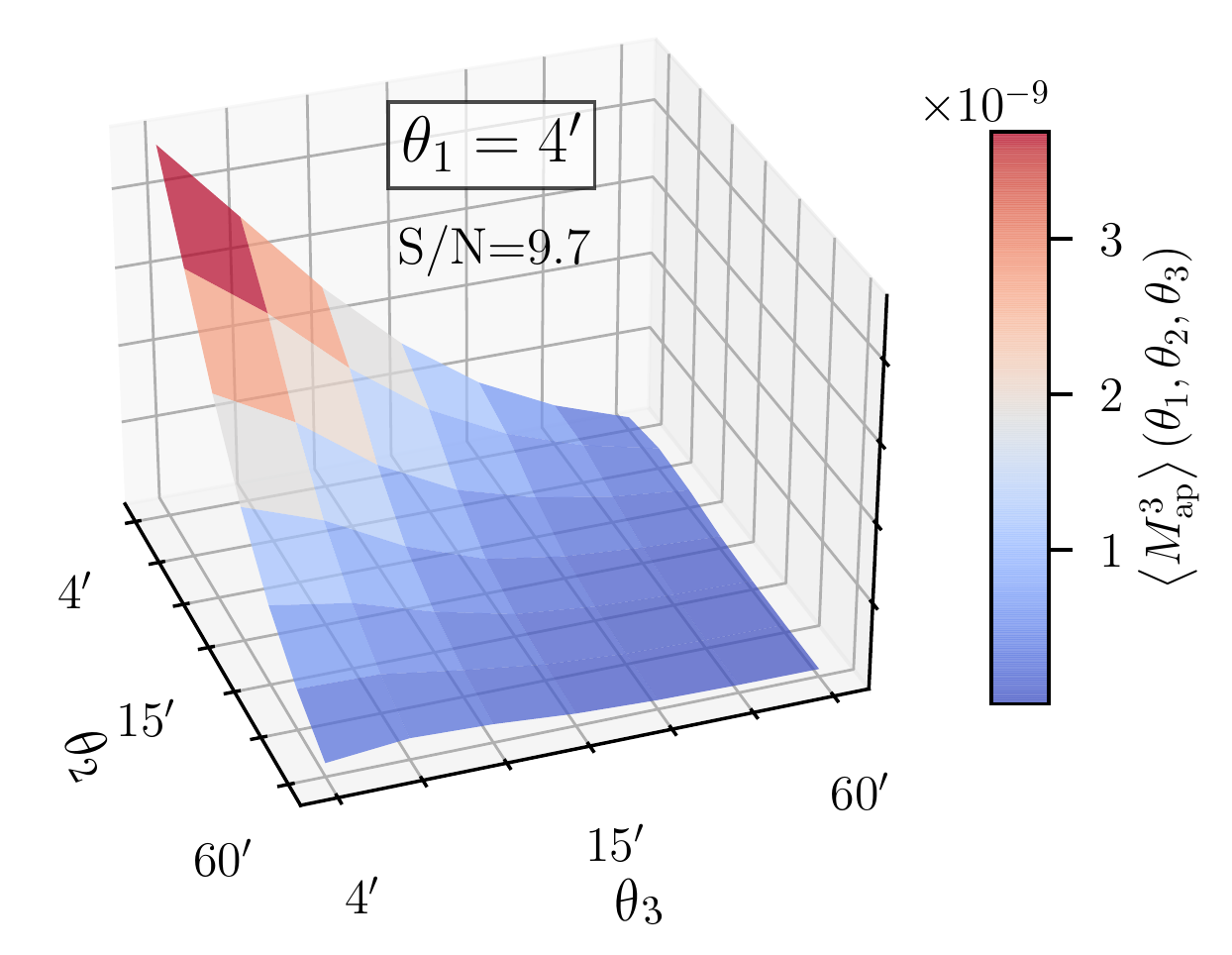}
	\includegraphics[width=1.0\columnwidth]{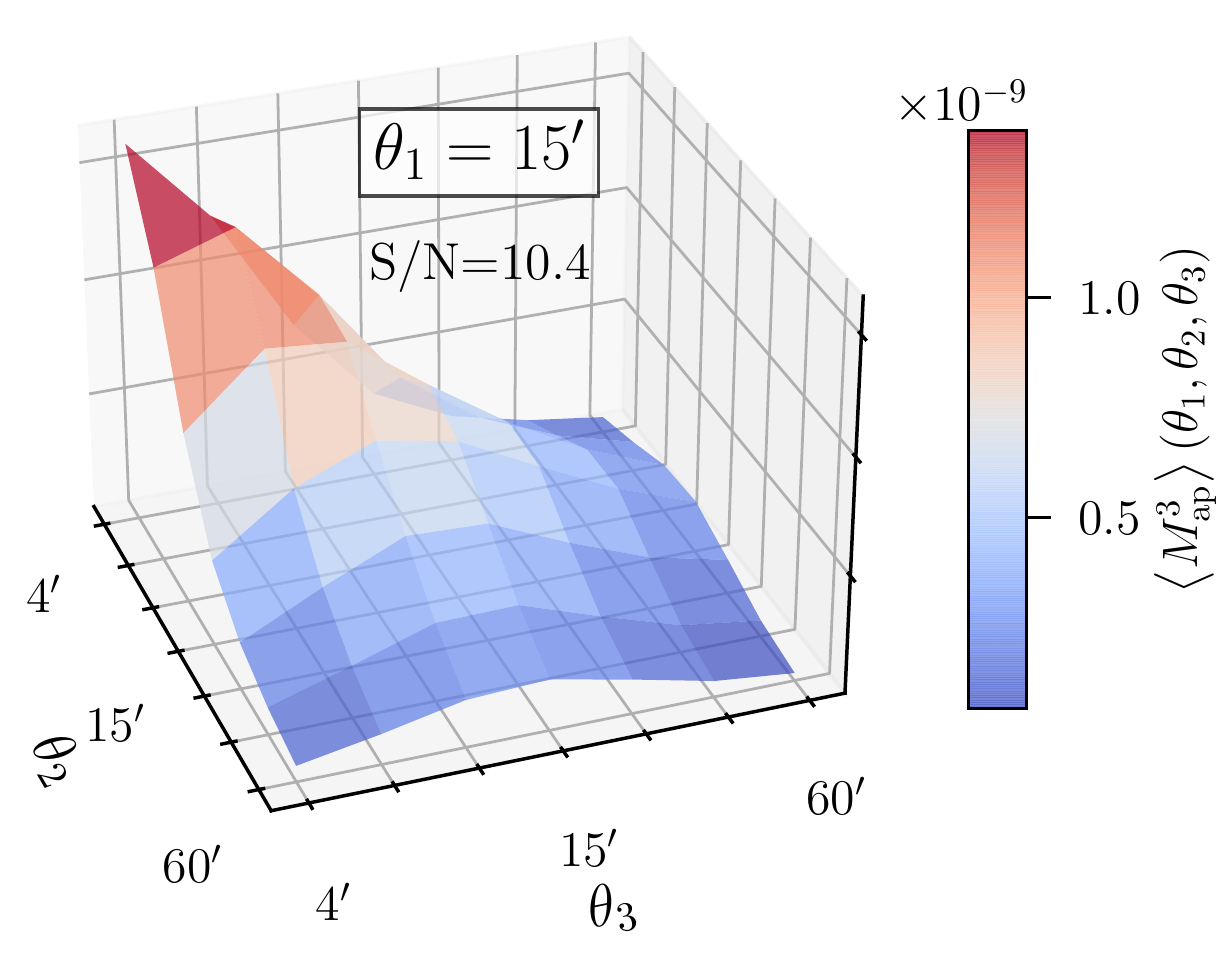}
	\includegraphics[width=1.0\columnwidth]{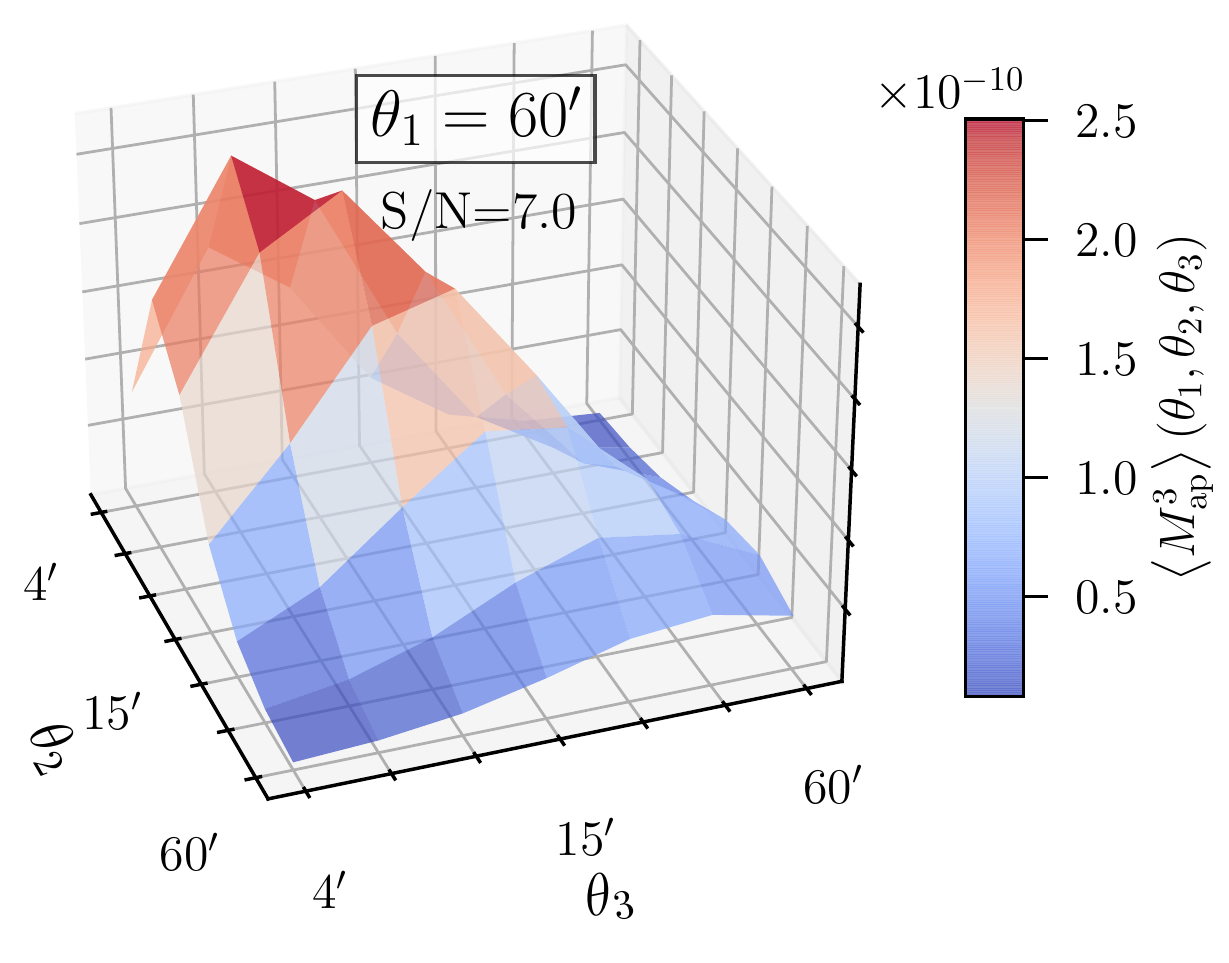}
    \caption{A detection of the generalized mass aperture skewness, expected to contain the entire 3pt information in the lensing field, with a dependence on three angular separations ($\theta_1, \theta_2,\theta_3$). We fix three choices of representative scales $\theta_1$ at 4, 15 and 60 arcmin, respectively the top, middle and bottom panels. In all cases, we find a significant detection of the generalized signal.}
    \label{fig: generalized Map3}
\end{figure} 

While the mass apertures involve a sum over many triangles and effectively mix their contributions to the signal, eqs.~(\ref{eq: Gammas in terms of projections})-(\ref{eq: Gamma_end}) on the other hand suggest that the natural components $\Gamma_i$ can be combined to separate specific triangle configurations and projections. Several triangle geometries were used by \cite{Takada_Jain_2002} to demonstrate that certain configurations (e.g., equilateral and isosceles triangles) have vanishing projections due to parity conservation, and that for general triangle shapes all 8 possible projections of $\gamma_{abc}$ with $a,b,c\in[t,\times]$ are non-zero. 

We can similarly explore the dependence of the signal on projection and configuration in our data by constructing $\gamma_{ttt}$ and $\gamma_{\times\times\times}$, the components with all shears projected tangentially and at 45$^o$ with respect to the triangle center respectively, via
\begin{equation}
    \gamma_{ttt}  =\frac{1}{4}\mathcal{R}\left[\Gamma_{0}+\Gamma_{1}+\Gamma_{2}+\Gamma_{3}\right]
\end{equation}
\begin{equation}
\gamma_{\times\times\times}  =\frac{1}{4}\mathcal{I}\left[-\Gamma_{0}+\Gamma_{1}+\Gamma_{2}+\Gamma_{3}\right],
\end{equation}
where $\mathcal{R}$ and $\mathcal{I}$ correspond to real/imaginary parts. Using the triangle side lengths ($d_1$, $d_2$, $d_3$) we obtain the shear signal for two types of configurations: isosceles triangles ($d_1\approx d_2\neq d_3$, with $\phi$ being the opening angle between $d_1$ and $d_2$), and equilateral triangles ($d_1\approx d_2 \approx d_3$)\footnote{These relations are only approximate in the data. For these specific configuration tests, we allow for small departures from exact triangle shapes, with side ratios binned with a $\pm 15\%$ tolerance in relative side lengths.}. Furthermore, we can separate ``small scale'' isosceles triangles with sides $d_1\approx d_2$ smaller than 5 arcmin, and ``large scale'' isosceles with $5 < d_1\approx d_2 < 60$ arcmin. We show our results in Fig. \ref{fig: configuration dependence}, where again dashed lines correspond to a measurement on a T17 mock. The left panel of the figure shows a characteristic oscillatory dependence on opening angle, somewhat similar to what was predicted for even-parity modes in \cite{Takada_Jain_2002} using a halo model approach, and in qualitative agreement with the T17 simulation result.  The right panel of Fig. \ref{fig: configuration dependence} shows the tangential and cross components of equilateral triangles as a function of angular separation $\theta_\mathrm{medium}=d_2\approx d_1 \approx d_3$. We find a significant signal in the even-parity $\gamma_{ttt}$ part, while the parity-violating term $\gamma_{\times\times\times}$  is consistent with zero; both are thus consistent with expectation. While the similarity of our signals with halo model studies such as \cite{Takada_Jain_2002}, \cite{Zaldarriaga_Scoccimarro2003} and \cite{Ho_White_2004} is visually striking, it is not exact. In particular, we find peaked signals on isosceles opening angles $\phi\to 0^o$ and $\phi\to 180^o$ that do not exactly match the expectation based on either work, but follow closely the T17 result. We believe that a quantitative comparison of these measured signals with theory and the information this could provide on gravity, nonlinear structure evolution and halo shapes certainly merits further exploration.

We further explore the general definition of the mass aperture skewness for three different aperture radii $\left\langle M^3_\mathrm{ap}\right\rangle(\theta_1 , \theta_2, \theta_3)$ in equation (\ref{eq: Map3 in terms of bispectrum}), and obtain the signal in some specific setups as shown in Fig. \ref{fig: generalized Map3}. We fix the aperture radius $\theta_1$ at 4, 15 and 60 arcmin, representing roughly the smallest, intermediate and largest scales probed with this statistic, and plot the signal as a function of the two other apertures. We find that the amplitude of the third-order mass aperture tends to be higher as we go to smaller scales. We note also that, while the generalized $\left\langle M^3_\mathrm{ap}\right\rangle(\theta_1 , \theta_2, \theta_3)$ contain the entire $E$-mode information of the field, they do not necessarily contain the highest signal-to-noise individually, a factor that should be taken into account in a future likelihood inference study. Nevertheless, in all cases we again find a significant detection of this particular lensing signal.

\subsection{Tomography}\label{sec: tomographic map3}

Motivated by the significant detections obtained in the non-tomographic regime, we proceed to split the DES Y3 catalog into redshift bins and attempt a first tomographic measurement of the third moment of the mass aperture.

We implement the same 2-bin redshift split described in Sec. \ref{sec: des data} on the T17 mock described in Sec. \ref{sec: takahashi}. The original, 4-bin redshift distributions in that mock resemble the actual DES Y3 $n(z)$'s but do not reproduce their substructure exactly, so we expect that 3pt statistics obtained from the mock should provide an approximate expectation for the scale dependence and amplitude of the tomographic signal on the data.

We present our results for the cross-tomographic mass apertures in the right-hand panel of Fig. \ref{fig: 1-aperture Map}, in qualitative agreement with the T17 result at most scales and redshift bins. We compute the signal-to-noise ratios $S/N$ again using eq. (\ref{eq: SN definition}), and find significant detections of cross-correlations of $\left\langle M^3_\mathrm{ap} \right\rangle$ that include the high-redshift bin $z_2$. For the complete  data vector built with the 4 concatenated cross-tomographic measurements and including their cross-covariances, we find a total $S/N$ of $15.0\sigma$. Interestingly, this detection is non-zero on angular scales that are relatively large ($\theta\sim 1^o$), reaching quasi-linear and linear regimes. This implies that non-Gaussian signals may add significant information to common two-point analyses even if these mostly rely on the linear regime due to conservative scale cuts (see, for instance, \cite{Gatti2021moments}).

We note several points related to this tomographic measurement. First, the signal in the higher redshift bin $z_2$ (red curve in the right panel of Fig. \ref{fig: 1-aperture Map}) is significantly larger than that for the lower bin $z_1$. As with the 2pt shear measurement, this trend can be attributed to the fact that the lensing kernel for the higher redshift bin probes more large-scale structure than the kernel limited to low redshifts. Second, the signal-to-noise of $\left\langle M_\mathrm{ap}(z_1)M_\mathrm{ap}(z_2)^2 \right\rangle$ ($S/N=13.3$) and $\left\langle M_\mathrm{ap}(z_2)^3 \right\rangle$ ($S/N=11.5$) are both higher than the non-tomographic case ($S/N=11.2$). While this may seem counter-intuitive at first, it is not against expectations: there are many low-redshift galaxy triplets in the non-tomographic sample whose 3pt correlations add significant noise but insignificant signal due to the lack of depth of the lensing kernel in the lowest redshift bin, and the overall $S/N$ goes up once these are removed.  Third, it is expected that $\left\langle M_\mathrm{ap}(z_1)M_\mathrm{ap}(z_2)^2 \right\rangle$ should have the highest $S/N$: for redshift bins with approximately the same number of galaxies, a cross-correlation contains a larger number of galaxy triplets than any auto-correlation, and additionally shot-noise contributions to the uncertainties are diagonal on the redshift bins.  

\begin{figure}
	\includegraphics[width=\columnwidth]{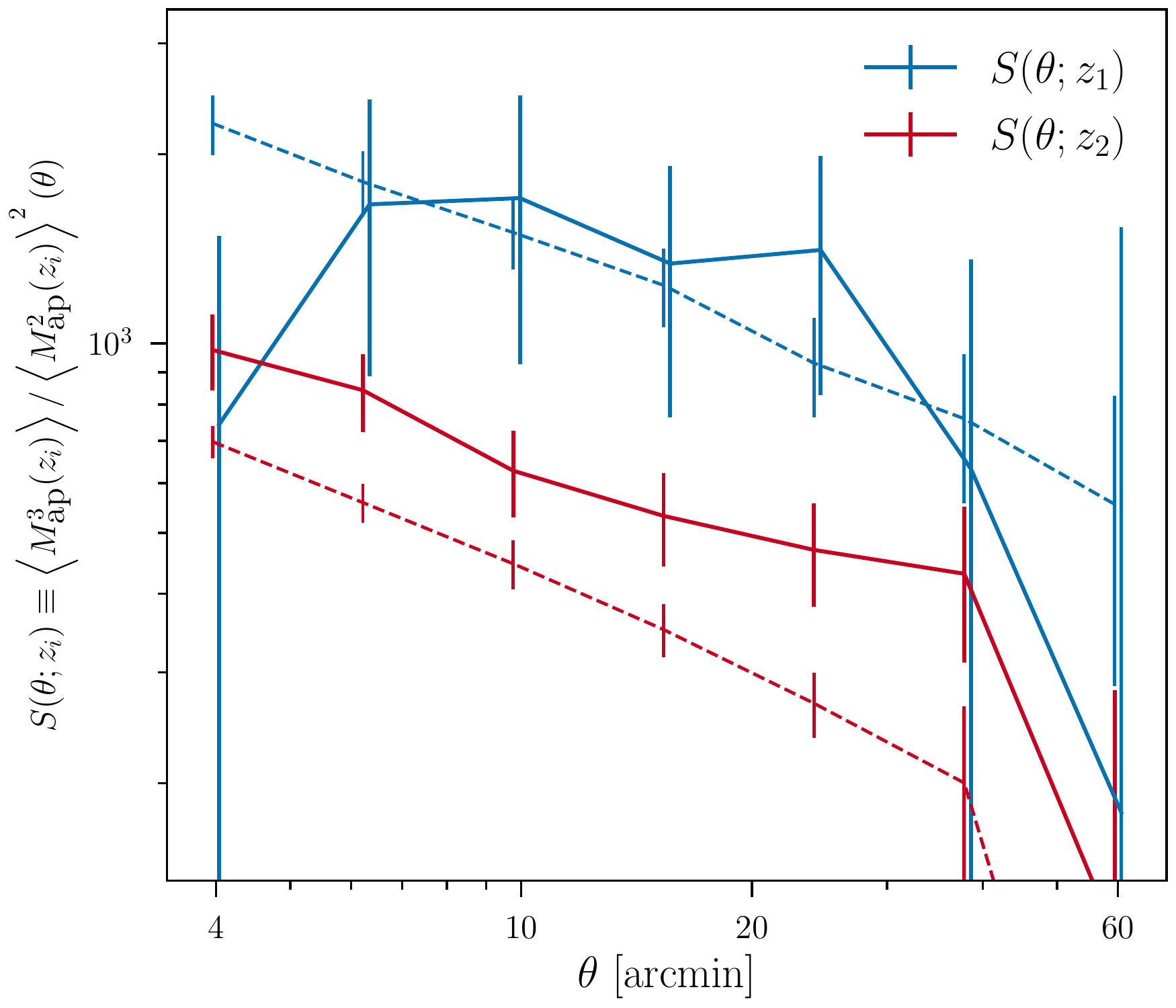}
    \caption{A measurement of the reduced mass aperture defined in eq. (\ref{eq: reduced map}). Solid lines and error bars correspond to measurements on the DES Y3 data  split into 2 wide redshift bins, and dashed lines correspond to measurements on an $N$-body mock based on the T17 simulations (see Sec. \ref{sec: takahashi}), which are not expected to fit the data, but serve as a guiding comparison. The redshift evolution  of the skewness parameter $S(\theta;z)$ indicates, according to expectation, that more non-Gaussian structure contributes to the signal at low-$z$ than at high-$z$. }
    \label{fig: Qs}
\end{figure}

In addition to the signals presented above, we measure the reduced skewness parameter in eq. (\ref{eq: reduced map}). We again use \textsc{TreeCorr} in order to estimate $\left\langle M^2_\mathrm{ap}\right\rangle$ in our data and mocks over the same patches where the 3pt observables were obtained. We show $S(\theta;z)$ in Fig. \ref{fig: Qs}. A significant redshift evolution of the reduced skewness parameter can be seen, with the low-$z$ bin showing more power than the high-$z$ bin. 
This is in line with our expectation that the shear field should be more Gaussian at higher redshift. This is due to the larger projection distance for high redshift, which means more uncorrelated structure contributes to the lensing and a version of the central limit theorem (considering the accumulated signal as a random walk along the line-of-sight)  makes the resulting shear field closer to Gaussian \citep{Bernardeau97,Jain97}. Note that there is no such expectation for the 3-dimensional density field, where the skewness is redshift-independent in leading order perturbation theory. The lensing skewness is largely independent of the power spectrum shape and normalization, and its approximate redshift evolution was given by e.g. \cite{Bernardeau97} who obtained $S\sim z^{-1.35}$. While that scaling depends on the cosmological model and the assumptions on the source redshift distribution, we find it to be in qualitative agreement with our measurement: for a representative scale of 10' the ratio $S(\theta=10';z_1)/S(\theta=10';z_2)$ is about 2 to 3, with the mean of redshift bins $z_1$ and $z_2$ being at 0.42 and 0.81 (see Sec. \ref{sec: data}), roughly following the expected scaling.   

\section{Assessment of Systematics}\label{sec: validation}

We now turn to the validation of the signal with the aim of showing that the detection is not contaminated by systematics of observational/instrumental origin. The results of the tests detailed below indicate that the significant 3rd order lensing signals found in DES Y3 data are of astrophysical and gravitational origin.  

\subsection{Estimator Uncertainties}\label{sec: estimator uncertainties}

Potential uncertainties in the estimation of $\Gamma_i$'s and $\map$ have three different sources, two of them originating from approximations needed to bring the computational runtime to a reasonable level and one, much easier to mitigate, stemming from the mass aperture filtering. We begin by describing this last one, the filtering feature, which we have essentially mitigated in this work by employing angular scale cuts at the measurement level.

The filter defined in eq. (\ref{eq: crittenden filter}) decays quickly as a function of angular separation, and is small (but not negligibly so) at separations of a factor $\lambda$ of about 3$\times$ wider than the angular bin at which \map~ is evaluated. This non-localization of the filter implies that, for a measurement of \map~ at an angular separation $\hat{\theta}$, the integrals over $\Gamma_i$ have significant support over a range $[\hat{\theta}/\lambda,\, \hat{\theta}\lambda]$ where $\lambda$ is a characteristic scale of filter. We employ a factor $\lambda=4$ after empirically testing estimates of \map~ over different angular ranges and finding them to stabilize very well at that chosen width. This choice is similar to previous studies \citep{Fu_etal_2014} and  justifies our choice of scales of [$1'$, $240'$] for $\Gamma_i(\theta_\mathrm{medium})$ and [$4'$, $60'$] for the aperture radii in \map. 

The other two potential sources of estimator uncertainties that we have explored are a decreased binning accuracy w.r.t. analogous calculations of 2pt functions, and the jackknifing method utilized. Binning accuracy in \textsc{TreeCorr} is determined by code parameters \texttt{binslop} and \texttt{binsize}. Larger values of the former allow for larger errors when binning triangles by ratios of their side lengths (see App. \ref{appendix: coordinates}), and larger values of the latter imply coarser binning by triangle configuration. We empirically vary both on a reduced number of data patches to verify their impact on our measurement. First, we find that the recommended value of \texttt{binsize}=0.1 is sufficient for the integration over $\Gamma_i$ and yields a stable \map. Second, while runtime increases prohibitively with smaller \texttt{binslop}, we find that a value of \texttt{binslop}=1.0 makes computing time feasible and does not bias the correlation functions, although it increases the diagonal covariance of the measurement by around 15\%. With these choices, we find that the computing time for 1M objects in 28 2.4GHz CPUs on the \texttt{Midway2} cluster\footnote{\url{https://rcc.uchicago.edu/}} is around 300 minutes (still very expensive when compared to a timing of under 20 minutes for the 2pt $\xi_\pm(\theta)$ auto-correlation of 25M galaxies in one redshift bin of DES Y3 with \texttt{binslop}=0.0, which approximates brute-force pair-counting). 

Finally, there are the uncertainties associated with the jackknife method, which we employ for more efficient parallelization and to obtain an estimate of the covariance matrix. A source of uncertainty comes from triangles whose corners are not all located on the same patch, because these triangles are not included in the subsequent calculations. We run a feasibly short test on the full 100M-object catalog by focusing only on several angular scales of approximately equilateral triangles and find that splitting the full footprint into disjoint patches misses approximately 10\% of the nearly equilateral triangles with a side length of 200 arcmin. The missing triplets enhance the shot noise contribution in those large scales, but should not contribute a bias because there is no preferential shear projection that is missed due to the patch splitting.  

\subsection{B-modes and Parity}

In general, a three-point signature of $B$-modes of astrophysical original can come from a limited number of effects. In particular, at the 3pt level the main sources of $B$-modes are intrisic alignments \citep{Semboloni_etal_2011,Troxel2012, Troxel_review} and the spatial clustering of source galaxies which are otherwise expected to randomly sample the survey footprint \citep{Schneider2002}. These effects are expected to be small compared to the lensing-induced $E$-mode signal, so at first a reasonable approach is to consider any significant $B$-mode detection as pointing to potential data systematics (PSF residuals, for instance).

Within the statistics we explore, the main correlations where $B$-modes could be searched for are $\left\langle M_\mathrm{ap}M^2_\times\right\rangle (\theta)$, which would point to $B$-modes correlated with $E$-modes. In the non-tomographic case, as shown in the left panel of Fig. \ref{fig: 1-aperture Map}, we find that the signal-to-noise of $\left\langle M_\mathrm{ap}M^2_\times\right\rangle (\theta)$ is compatible with the null-hypothesis according to the definition in eq.~(\ref{eq: SN definition}), meaning $S/N$ is lesser than 1 or imaginary. In a similar way, we verify that $\left\langle M_\mathrm{ap}(z_1) M^2_\times (z_2)\right\rangle$ and  $\left\langle M_\mathrm{ap}(z_2) M^2_\times (z_1)\right\rangle$, the tomographic versions of the same test which would respectively point to $B$-modes in the higher(lower) redshift bin correlating with $E$-modes in the lower(higher) redshift bin, are also consistent with the ``Null'' condition defined in eq.~(\ref{eq: SN definition}). 

Other correlations including odd powers of the $B$-mode field $M_\times$ such as $\left\langle M^2_\mathrm{ap}M_\times\right\rangle (\theta)$ are expected to vanish due to parity \citep{Schneider2003}. A parity-violating field would necessarily come from systematics of the data, as no astrophysical source could produce it. We indeed find the parity-violating terms $\left\langle M^2_\mathrm{ap}M_\times\right\rangle (\theta)$ and $\left\langle M^3_\times\right\rangle (\theta)$ presented in the left panel of Fig. \ref{fig: 1-aperture Map} to be consistent with the null-hypothesis. Finally, we have also shown in Fig. \ref{fig: configuration dependence} another parity-violating correlation,  $\gamma_{\times\times\times}(\theta)$ for approximately equilateral triangles, which is similarly consistent with zero.

\subsection{PSF Residuals}

We follow the approach of \cite{rowe2010} in order to estimate the contribution of additive PSF modeling errors to our lensing observables. We obtain the mass aperture skewness of the so-called ``$\rho$-statistics'' (see Appendix \ref{appendix: rho stats}), which  quantify the residual correlations caused by errors in the PSF modeling and deconvolution, modulated by empirically-obtained coefficients $\alpha$ and $\beta$.

We estimate the PSF uncertainty impact via eq. (\ref{eq: Gamma PSF correction}) using a catalog of stars to compare them to the actual data signal. In doing so, we need input values for the coefficients $\alpha$ and $\beta$ that multiply deconvolution errors and modeling residuals, respectively. We set $\alpha=0.01$ and $\beta=2$ as inputs for the additive contaminations, considering the bounds on these parameters presented in \citet*{Gatti2021} (respectively $\alpha= 0.001 \pm 0.005$ and $\beta=1.09  \pm 0.07$). This choice of input values is a very conservative one, which amplifies the estimated impact of these systematics.  As the additive PSF contaminations considered here have their origin in the 1-point ellipticities, we do not expect the values of those coefficients to depend on which statistics are used to measure them (apart from practical aspects such as the signal-to-noise of the chosen statistic). We therefore do not pursue a measurement of $\alpha$ and $\beta$ based on 3pt observables, and utilize those bounds obtained in \citet*{Gatti2021} based on 1- and 2-point PSF correlations.

 Despite the conservative choice in input coefficients, we find additive PSF systematics to be entirely negligible. We show in  Fig. \ref{fig: PSF} a breakdown of the PSF contributions to individual skewness component ($\left\langle M^3_\textrm{ap} \right\rangle$, $\left\langle M^2_\textrm{ap} M_\times \right\rangle$, etc) and by PSF correlation type ($\left\langle e_p^3 \right\rangle$, $\left\langle e_p^2 q \right\rangle$, etc), where $e_p$ is the PSF ellipticity and $q$ the ellipticity residual error after modeling. In all cases, we find the 3rd order moments of PSF uncertainties to be negligible, well below a percent of the $E$-mode data signal $\left\langle M^3_\textrm{ap} \right\rangle$.


\subsection{Mean Shear and Other Observational Systematics}

Several other features of 3pt statistics are also relevant for their robustness against systematics. In particular we consider contributions to the signal arising from a residual mean shear in ellipticities $\left\langle e_1\right\rangle$ and $\left\langle e_2\right\rangle$.

While a mean shear that is coherent  across angular scales  produces a $\xi_+$ signal (eq. \ref{eq: xipm definition}) at the 2pt level, it does not produce any signature on the $\Gamma_i$. This can easily be demonstrated by considering a constant shear field in cartesian coordinates, $\gamma=\gamma_1 +i\gamma_2=c_1$, coherent across some angular length scale. For 2pt functions $\xi_\pm$ we project shears along the direction $\alpha+\psi$, where $\alpha$ is the direction of the line that connects the galaxy pair and $\psi$ is the (random) orientation of the pair with respect to the reference of the cartesian coordinates, so $\gamma\to\gamma'=\gamma\exp\left[-2i(\alpha+\psi)\right]$. Then the natural 2pt functions of the field are  $\xi_+=\left\langle \gamma' \gamma'^* \right\rangle = c_1^2$ and $\xi_-=\left\langle \gamma' \gamma' \right\rangle=c_{1}^{2}\left\langle \exp\left[-4i\left(\alpha+\psi\right)\right]\right\rangle=0$ as the averaging is essentially over the multiple random orientations $\psi$. 

For the natural 3pt functions, in comparison, the projection of each of the 3 shear components is along a different direction ($\alpha+\psi$, $\beta+\psi$ or $\delta+\psi$) and many reference points are possible - the triangle incenter, the center of the side opposing a given angle, etc \citep{Schneider2003_natural_components}, with some projections leading to $\alpha+\beta+\delta=0$. In the same situation of a constant shear in cartesian coordinates we have, for an example case: $\Gamma_{0}=\left\langle c_{1}^{3}\exp\left[-6i\left(\alpha+\beta+\delta\right)-6i\psi\right]\right\rangle=0$ due to the averaging over $\psi$, and similarly for all other $\Gamma_i$ with $i=1,2,3$ defined in eqs.~(\ref{eq: Gammas in terms of projections})-(\ref{eq: Gamma_end}). This insensitivity to an additive mean shear over coherent scales can be useful when compared to 2pt functions because it would not lead to the requirement of an extra correction at the data level as in (\citet*{Gatti2021}), and would potentially minimize the need for corrections due to additive systematics such as presented in \cite{Kitching2021}. 

We additionally expect that any other observational systematics that arise from statistics that are well described by Gaussian processes should have negligible contributions to 3pt functions. A potential example which we leave for a further exploration is the  atmospheric contribution to PSFs. As that is well characterized by Gaussian processes with vanishing odd-order correlations, we expect it to be significantly suppressed in importance when dealing with 3pt shear correlations.

\section{Comparison with Previous Work}

Among the several types of 3pt shear statistics presented so far, some had already been detected and explored in the survey science literature while others had not.  In what follows, we compare our findings with a number of previous results.

As a starting point, our 3pt $S/N$ can be compared with the 2pt DES Y3 cosmic shear measurements. The null-hypothesis signal-to-noise defined in  eq. (\ref{eq: SN definition}) yields $S/N=40.2$ for the joint $\xi^{ij}_\pm$ data vector (eq. \ref{eq: xipm definition}) presented in \cite{Amon21}, \citet*{Secco21}\footnote{Note that the definition of $S/N$ utilized in these works is different than the one employed here.} \textit{before} the removal of relatively small angular scales that are not included in the likelihood due to modeling uncertainties (a total $N_\mathrm{d.o.f.}=400$ degrees of freedom). After ``fiducial'' scale cuts, the DES Y3 cosmic shear data vector has $S/N=27.5$ ($N_\mathrm{d.o.f}=227$), and after ``optimized'' scale cuts we obtain  $S/N=30.1$ ($N_\mathrm{d.o.f}=273$). 

While the signal-to-noise ratio of our 3pt measurements are smaller than the corresponding 2pt S/N, it is realistic to expect that real-space 3pt shear correlations can tighten posteriors in key cosmology results because parameter degeneracies are different between two- and three-point functions. That is indeed the case with \cite{Gatti2021moments}, wherein an improvement of $\sim 15\%$ is seen in the lensing amplitude $S_8$ when combining second and third order moments of the lensing convergence. 

Regarding three-point detections of cosmic shear observables, Stage-II surveys presented some of the first results: a first detection was claimed by \cite{Bernardeau2002}
in the VIRMOS-DESCART 8.5deg$^2$ survey \citep{Waerbekeetal2002}, followed by
detections of the third moment of the mass aperture 
 by \cite{JBJ04} with the CTIO 75deg$^2$ survey data, \cite{Semboloni_etal_2011} 
 with HST COSMOS data \citep{COSMOS, Schrabback2010} and, more recently, \cite{Fu_etal_2014} with CFHTLenS data \citep{Erben2013}. These first detections of lensing third moments with signal-to-noise around $3\sigma$ advanced the field. Our measurements significantly improve upon those  detections and bring them up to $S/N$ of around $15\sigma$, a significance that enables quantitative interpretation.

To the best of our knowledge and at the time of this writing, we have reported in this work the first significant detection of the four natural 3pt cosmic shear components (Fig. \ref{fig: non-tomographic}), the first detection of tomographic 3pt mass aperture signals (right panel of Fig. \ref{fig: 1-aperture Map}), and the first significant detection of components split by their configuration dependence (Fig. \ref{fig: configuration dependence}). Equally important, our measurement (along with the \cite{Gatti2021moments} measurement of the skewness of $\kappa$ in the same data),  extends to large scales approaching 1 degree, where quasilinear theory is reliable and uncertainties due to baryonic physics can be neglected. Thus it will enable robust interpretations of cosmology and gravitational physics.

\begin{figure*}
	\includegraphics[width=1.5\columnwidth]{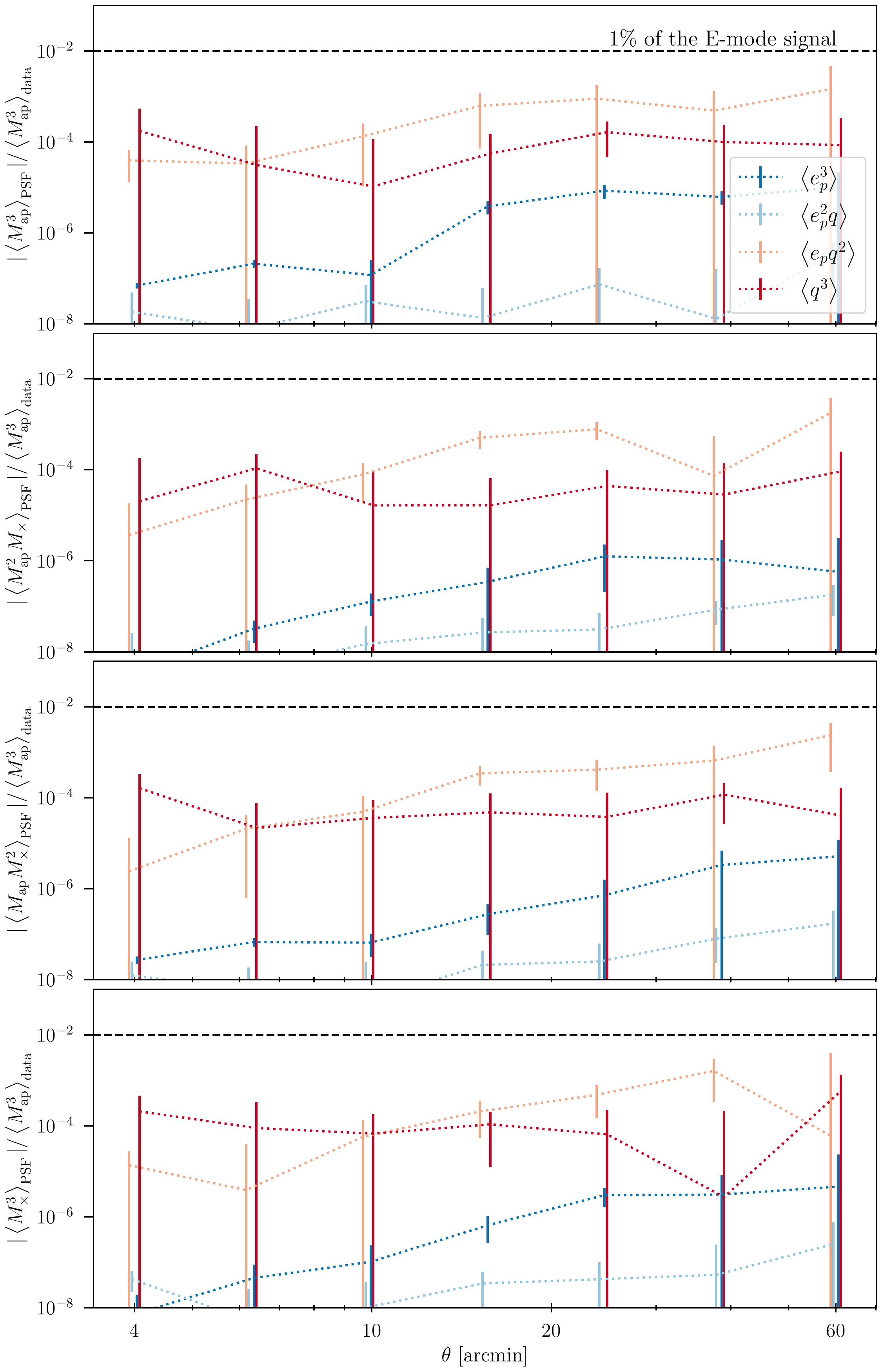}
    \caption{Upper bound on PSF systematics due to their potentially incorrect modelling and deconvolution, assuming coefficients $\alpha=0.01$ and $\beta=2.0$ in eq. (\ref{eq: Gamma PSF correction}).  Horizontal axes show angular separations and vertical axes in each panel, from top to bottom respectively, show the absolute value of PSF $\left\langle M^3_\textrm{ap} \right\rangle$, $\left\langle M^2_\textrm{ap} M_\times \right\rangle$, $\left\langle M_\textrm{ap} M^2_\times \right\rangle$ and $\left\langle M^3_\times \right\rangle$ correlations divided by the $E$-mode signal $\left\langle M^3_\textrm{ap} \right\rangle$ of the data, which is always at the sub-percent level and significantly smaller than the measurement errors. The dashed black line shows the 1\% level and blue, light blue, salmon and red lines correspond to different cross-correlation of PSF properties $e_p$ (the PSF ellipticity) and $q$ (the ellipticity residual error after modeling). }
    \label{fig: PSF}
\end{figure*}

\section{Conclusions and Outlook}\label{sec: conclusion}

Using over 100M galaxies spread across the 4,143deg$^2$ footprint of the first 3 years of data from the Dark Energy Survey, we presented measurements of the three-point correlations of the lensing shear field. We also combined all three point correlations into the third moment of the mass aperture statistic and verified that systematics of observational origin are negligible in our measurements.  We expect this work to be a stepping stone for future applications of these 3pt statistics, in particular a joint 2pt and 3pt cosmology analyses. Our main results are summarized below:

\begin{itemize}
    \item In a non-tomographic analysis, we measure the natural cosmic shear correlations $\Gamma_i$ (the 3pt functions analogous to the two-point functions $\xi_\pm$) in DES Y3 data at high signal-to-noise ($2.5\lesssim S/N\lesssim 7.0$ for the real part of the correlations) and also explore the triangle configuration dependence of 3pt shear projections (respectively Figs. \ref{fig: non-tomographic} and \ref{fig: configuration dependence});
    
    \item Also in a non-tomographic setting, we measure the skewness of the mass aperture statistic \map~ both in 1 aperture radius and in the generalized case of 3 aperture radii (left panel of Fig. \ref{fig: 1-aperture Map} and Fig. \ref{fig: generalized Map3}). The detection significance in all cases is very high ($7.0\lesssim S/N\lesssim 11.0$);
    
    \item We detect, for the first time, a tomographic \map($\theta$) signature with high significance (total tomographic $S/N=15.0$) and additionally verify an expected redshift evolution of the skewness parameter $S(\theta;z)$ (respectively the right panel of Fig. \ref{fig: 1-aperture Map} and Fig. \ref{fig: Qs});
    
    \item We verify that the third-order signatures found are robust against $B$-mode systematics, parity-violating contributions and PSF modeling errors, thus validating that our measurements are likely a result of astrophysical and gravitational phenomena (left panel of Fig. \ref{fig: 1-aperture Map}, right panel of Fig. \ref{fig: configuration dependence} and Fig. \ref{fig: PSF});
    
    \item We reproduce the main results in an N-body mock catalog and verify that  overall angular scale dependences and signal amplitudes of our measurements are broadly consistent with theoretical expectations.
\end{itemize}

Given the high $S/N$ of the data vectors here presented and the fact that systematics of observational origin are well under control, carrying forward with a cosmological analysis is a reasonable path. It is also interesting to note that our detected signals are non-zero even on relatively large angular separations of a degree or more, implying that non-Gaussian information coming from quasi-linear and linear scales could significantly add even to a conservative 2pt cosmic shear analysis. We do, however, identify below several challenges that a joint 2pt+3pt program would face. 

First, analytic covariances for higher order moments of shear are notoriously complex, and their uncertainties can significantly affect parameter posteriors. While it remains to be tested, it is possible that the jackknife approach employed here for the simple $S/N$ estimates might not be sufficiently accurate for the more subtle inference of cosmological parameters. A mock-based covariance would be straightforward method, but we note that the 3pt measurement runtime is computationally expensive and could make that approach impractical unless we select 3pt statistics that minimize that computational cost.
In particular, \map~ and its generalized form have a high signal-to-noise with a relatively small number of data points, which would make the use of mock covariances more feasible.      

Second, the modeling of astrophysical systematics such as intrinsic alignments and baryons is likely to preclude the use of small angular scales presented in our measurements, and therefore it might be necessary to remove part of those data points when fitting a theory model. However, the statistical uncertainties in our measurements are large compared to those for 2pt cosmic shear, so it is not unreasonable to expect that in fact relatively simple theory modeling can be used for the 3pt data vector and still maintain an acceptable level of potential biases. We also point out that the nonlinear dark matter bispectrum modeling itself is a challenge, although methods based on fitting formulas calibrated against simulations have been employed in the literature \citep{Lazanu2016,bihalofit}.

Third, redshift and shape measurement uncertainties  propagate significantly into 3pt observables. These uncertainties are calibrated to high accuracy and precision in 2pt analyses, and a comparably careful analysis is needed for 3pt correlations to determine their contribution to the error budget.

Many of the challenges above have already been addressed in cosmology studies including higher order lensing correlations, in particular in the convergence moments work of \cite{Gatti2021moments}. In detail, the covariance matrix estimation was made feasible in that work with a data compression technique retaining a number of data points smaller than the full length of the data vector. Also, scale cuts were determined by the impact of baryonic physics and other astrophysical contaminants (such as intrinsic alignments and 3rd order contributions such as source clustering) were modeled or shown to be negligible. Finally, the nonlinear matter bispectrum was obtained with a fitting formula calibrated on simulations \citep{SC01}, and its computation was made faster with an emulator technique.  

We expect to employ similar methodologies for the real space analysis of 2pt+3pt cosmic shear, but with some differences in the details owing to the different choice of estimators.  The resulting constraints would provide an important consistency check to the results of \cite{Gatti2021moments}, with the advantage that the real space statistics presented in this work (specifically the general, three-aperture radii $\left\langle M^3_\textrm{ap}\right\rangle (\theta_1, \theta_2, \theta_3)$) are guaranteed to contain the total $E$-mode content in the shear field along with specific configuration-dependent information. Finally, our measurements of the full 3-point function of the shear field lay the groundwork to test for primordial non-Gaussianity in the density field, e.g. via constraints on the $f_{\mathrm{NL}}$ parameter, as studied theoretically by \cite{Takada_Jain_2003_correlations} and \cite{Hilbert2012}. 

With many practical challenges overcome and a steadily increasing level of maturity, it is realistic to expect that that lensing 2pt+3pt analyses will be among the central probes of $S_8$ and the Dark Energy equation-of-state parameter $w$ in current and future surveys such as the Vera C. Rubin Observatory's Legacy Survey of Space and Time\footnote{\url{https://www.lsst.org}} (LSST), ESA's Euclid mission\footnote{\url{https://www.euclid-ec.org}} and the Roman Space Telescope\footnote{\url{https://roman.gsfc.nasa.gov}}. That is especially important since these experiments represent a massive investment of resources, and extracting as much useful information as possible from their data is highly desirable.

\begin{acknowledgments}
We would like to thank Lucas Porth for comments and Ryuichi Takahashi and collaborators for making the T17 simulations utilized in this work publicly available. This work was completed in part with resources provided by the University of Chicago Research Computing Center. MJ is supported in part by National Science Foundation Award 1907610. BJ is supported in part by the US Department of Energy grant DE-SC0007901. CC is supported by DOE grant DE-SC0021949.

Funding for the DES Projects has been provided by the U.S. Department of Energy, the U.S. National Science Foundation, the Ministry of Science and Education of Spain, 
the Science and Technology Facilities Council of the United Kingdom, the Higher Education Funding Council for England, the National Center for Supercomputing 
Applications at the University of Illinois at Urbana-Champaign, the Kavli Institute of Cosmological Physics at the University of Chicago, 
the Center for Cosmology and Astro-Particle Physics at the Ohio State University,
the Mitchell Institute for Fundamental Physics and Astronomy at Texas A\&M University, Financiadora de Estudos e Projetos, 
Funda{\c c}{\~a}o Carlos Chagas Filho de Amparo {\`a} Pesquisa do Estado do Rio de Janeiro, Conselho Nacional de Desenvolvimento Cient{\'i}fico e Tecnol{\'o}gico and 
the Minist{\'e}rio da Ci{\^e}ncia, Tecnologia e Inova{\c c}{\~a}o, the Deutsche Forschungsgemeinschaft and the Collaborating Institutions in the Dark Energy Survey. 

The Collaborating Institutions are Argonne National Laboratory, the University of California at Santa Cruz, the University of Cambridge, Centro de Investigaciones Energ{\'e}ticas, 
Medioambientales y Tecnol{\'o}gicas-Madrid, the University of Chicago, University College London, the DES-Brazil Consortium, the University of Edinburgh, 
the Eidgen{\"o}ssische Technische Hochschule (ETH) Z{\"u}rich, 
Fermi National Accelerator Laboratory, the University of Illinois at Urbana-Champaign, the Institut de Ci{\`e}ncies de l'Espai (IEEC/CSIC), 
the Institut de F{\'i}sica d'Altes Energies, Lawrence Berkeley National Laboratory, the Ludwig-Maximilians Universit{\"a}t M{\"u}nchen and the associated Excellence Cluster Universe, 
the University of Michigan, the National Optical Astronomy Observatory, the University of Nottingham, The Ohio State University, the University of Pennsylvania, the University of Portsmouth, 
SLAC National Accelerator Laboratory, Stanford University, the University of Sussex, Texas A\&M University, and the OzDES Membership Consortium.

The DES data management system is supported by the National Science Foundation under Grant Numbers AST-1138766 and AST-1536171. The DES participants from Spanish institutions are partially supported by MICINN under grants ESP2017-89838, PGC2018-094773, PGC2018-102021, SEV-2016-0588, SEV-2016-0597, and MDM-2015-0509, some of which include ERDF funds from the European Union. IFAE is partially funded by the CERCA program of the Generalitat de Catalunya. Research leading to these results has received funding from the European ResearchCouncil under the European Union’s Seventh Framework Program (FP7/2007-2013) including ERC grant agreements 240672, 291329, and 306478. We acknowledge support from the Brazilian Instituto Nacional de Ciência e Tecnologia (INCT) do e-Universo (CNPq grant 465376/2014-2). We  acknowledge support from the Australian Research Council Centre of Excellence for All-sky Astrophysics (CAASTRO), through project number CE110001020.

This manuscript has been authored by Fermi Research Alliance, LLC under Contract No. DE-AC02-07CH11359 with the U.S. Department of Energy, Office of Science, Office of High Energy Physics. The United States Government retains and the publisher, by accepting the article for publication, acknowledges that the United States Government retains a non-exclusive, paid-up, irrevocable, world-wide license to publish or reproduce the published form of this manuscript, or allow others to do so, for United States Government purposes.

Based in part on observations at Cerro Tololo Inter-American Observatory, 
National Optical Astronomy Observatory, which is operated by the Association of 
Universities for Research in Astronomy (AURA) under a cooperative agreement with the National 
Science Foundation.
This work made use of Matplotlib \citep{matplotlib} and NASA's Astrophysics Data System Bibliographic Services (ADS).

\end{acknowledgments}

\appendix

\section{Derivation of PSF corrections}\label{appendix: rho stats}

As ellipticities $e$ are measured from galaxy images, their PSF must be deconvolved. As in \citet*{Gatti2021}, we define the errors in the PSF modeling as well improper deconvolution both as additive contributions to the measured ellipticities:
\begin{equation}\label{eq:additive_psf_correction}
e =\gamma+\delta_{e},
\end{equation}
where the additive factor $\delta_e$ is defined by 
\begin{equation}\label{eq: alpha and beta}
\delta_e\equiv \alpha e^p + \beta q; \quad\quad q\equiv e_* - e^p,
\end{equation}
where $e^{p}$ is the modeled PSF elipticity (referred to as $e_\textrm{model}$ in \citet*{Gatti2021}) and $e_*$ is the actually measured PSF. That means the coefficients $\alpha$ and $\beta$ are respectively interpreted as a leakage of the modeled PSF shape onto the galaxy ellipticity $e$ (coming possibly from incorrect deconvolutions) and errors in the interpolation of the PSF shape.  The shear fields $e^p$ and $q$ are estimated from \textit{reserved stars} which do not contribute to the PSF fitting, that is, where both the modeled PSF and the true PSF are known, otherwise we would have $q\to0$ by construction.  

Using the same definitions in eq. (\ref{eq: M definition}), we can propagate the PSF correction in eq. (\ref{eq:additive_psf_correction}) to the 1-point quantities:
\begin{align*}
M_{\textrm{ap}}(R) & =\int d^{2}R\,Q(R)\gamma_{t}+\int d^{2}R\,Q(R)\left[\alpha e_{t}^{p}+\beta q_{t}\right]\\
M_{\times}(R) & =\underset{=0}{\underbrace{\int d^{2}R\,Q(R)\gamma_{\times}}}+\int d^{2}R\,Q(R)\left[\alpha e_{\times}^{p}+\beta q_{\times}\right].
\end{align*}
 As the cross-projections of the PSF residuals $e^p_\times$ and $q_\times$ can generally have non-zero statistical moments, we see that the additive PSF errors defined above can contaminate both E-modes and B-modes. As a stepping stone for the third-order case, we can again follow \cite{JBJ04} and get, for the second-order mass aperture:
\begin{align*}
\left\langle M^{2}\right\rangle  & =\int d^{2}R_{1}d^{2}R_{2}\,Q(R_{1})Q(R_{2})\left\langle \left(\gamma+\delta_{e}\right)\left(\gamma+\delta_{e}\right)\right\rangle \\
 & \quad\times\exp\left(-2i\left(\phi_{1}+\phi_{2}\right)\right)\\
\left\langle MM^{*}\right\rangle  & =\int d^{2}R_{1}d^{2}R_{2}\,Q(R_{1})Q(R_{2})\left\langle \left(\gamma+\delta_{e}\right)\left(\gamma+\delta_{e}\right)^{*}\right\rangle \\
 & \quad\times\exp\left(-2i\left(\phi_{1}-\phi_{2}\right)\right).
\end{align*}
We can safely assume that the expected value of correlations between the gravitational shear and PSF residuals is zero so the cross-terms $\left\langle \gamma\delta_{e}\right\rangle $ vanish. Then, defining the PSF correlations $\xi_{\pm}^{\textrm{psf}}$ analogously to how the (gravitational) shear correlations are defined, that is $\xi_{+}\equiv\left\langle \gamma\gamma^{*}\right\rangle$, $\xi_{-}\equiv\left\langle \gamma\gamma\exp(-4i\theta)\right\rangle$,
we see that the PSF corrections are simply additive at the mass aperture level: 
\begin{align}
\left\langle M^{2}\right\rangle (R) & =\int\frac{s\,ds}{R^{2}}\left(\xi_{-}^{\gamma}(s)+\xi_{-}^{\textrm{psf}}(s)\right)T_{-}\left(\frac{s}{R}\right)\\
\left\langle MM^{*}\right\rangle (R) & =\int\frac{s\,ds}{R^{2}}\left(\xi_{+}^{\gamma}(s)+\xi_{+}^{\textrm{psf}}(s)\right)T_{+}\left(\frac{s}{R}\right)
\end{align}
where $\xi_{\pm}^{\textrm{psf}}\equiv\alpha^{2}\left\langle e_{p}e_{p}\right\rangle _{\pm}+\alpha\beta\left\langle e_{p}q\right\rangle _{\pm}+\beta^{2}\left\langle qq\right\rangle _{\pm}$, and where the functions $T_\pm$ are defined in Appendix \ref{appendix: coordinates}.

 The terms $\left\langle M_{\mathrm{ap}}^{2}\right\rangle $  and $\left\langle M_{\times}^{2}\right\rangle$ can be expressed as simple linear combinations of the quantities above \citep{JBJ04}. While $\left\langle M_{\times}^{2}\right\rangle$ would represent B-mode signal which can generally become non-zero in the presence of uncorrected PSF errors, the term $\left\langle M_{\textrm{ap}}M_{\times}\right\rangle$, if found to be non-negligible, would additionally imply a parity-violating contribution.

The reasoning above also applies to the third-order moments of the same observables. We define the 3pt PSF correlations in the same way we define the natural components of the shear signal
and write 
\begin{align}
\left\langle M^{3}\right\rangle (R) & =-\int d^{2}R_{1}d^{2}R_{2}d^{2}R_{3}\,Q(R_{1})Q(R_{2})Q(R_{3})\nonumber \\
 & \quad\times\left\langle \left(\gamma+\delta_{e}\right)^{3}\exp\left[-2i(\alpha+\beta+\delta)\right])\right\rangle \nonumber  \\
 & =-\int\frac{s\,ds}{R^{2}}\int\frac{d^{2}t}{2\pi R^{2}}\left(\Gamma_{0}+\Gamma_{0}^{\textrm{psf}}\right)T_{0}(s,t)\label{eq: M3 PSF}
\end{align}

\begin{align}
\left\langle M^{2}M^{*}\right\rangle (R) & =\int d^{2}R_{1}d^{2}R_{2}d^{2}R_{3}\,Q(R_{1})Q(R_{2})Q(R_{3})\nonumber \\
 & \quad\times\left\langle \left(\gamma+\delta_{e}\right)^{2}\left(\gamma+\delta_{e}\right)^{*}\exp\left[-2i(\alpha+\beta-\delta)\right]\right\rangle \nonumber \\
 & =\int\frac{s\,ds}{R^{2}}\int\frac{d^{2}t}{2\pi R^{2}}\left(\Gamma_{1}+\Gamma_{1}^{\textrm{psf}}\right)T_{1}(s,t)\label{eq: M2Mstar PSF}
\end{align}
where we have introduced the PSF correction at the 3pt level as
\begin{equation}
    \Gamma_{0,1}^{\textrm{psf}}=\alpha^{3}\left\langle e_{p}^3\right\rangle _{0,1}+3\alpha^{2}\beta\left\langle e_{p}^2q\right\rangle _{0,1}+3\alpha\beta^{2}\left\langle e_{p}q^2\right\rangle _{0,1}+\beta^{3}\left\langle q^3\right\rangle _{0,1}.\label{eq: Gamma PSF correction}
\end{equation}

The derivation above assumes that cross-terms of the type  $\left\langle \gamma\delta_{e}^{2}\right\rangle $
or $\left\langle \gamma^{2}\delta_{e}\right\rangle $ are null when averaged over large ensembles, as both of these terms boil down to whether the 1(2)-point gravitational shear correlates with the 2(1)-point PSF's, which should not be the case. The expressions for $\left\langle M_{\textrm{ap}}^{3}\right\rangle $,
$\left\langle M_{\textrm{ap}}^{2}M_{\times}\right\rangle $, $\left\langle M_{\textrm{ap}}M_{\times}^{2}\right\rangle $
and $\left\langle M_{\times}^{3}\right\rangle $ can be obtained from the ones above as shown in \cite{JBJ04}, and it remains true that $\left\langle M_{\textrm{ap}}^{2}M_{\times}\right\rangle $
and $\left\langle M_{\times}^{3}\right\rangle $ are null in order to conserve parity, while $\left\langle M_{\textrm{ap}}M_{\times}^{2}\right\rangle$ \textit{may} include non-zero PSF B-modes that correlate with E-modes (in addition to astrophysical B-mode contributions).

\section{Definition of coordinates and \textsc{TreeCorr} internal variables}\label{appendix: coordinates}

\begin{figure}
	\includegraphics[width=0.8\columnwidth]{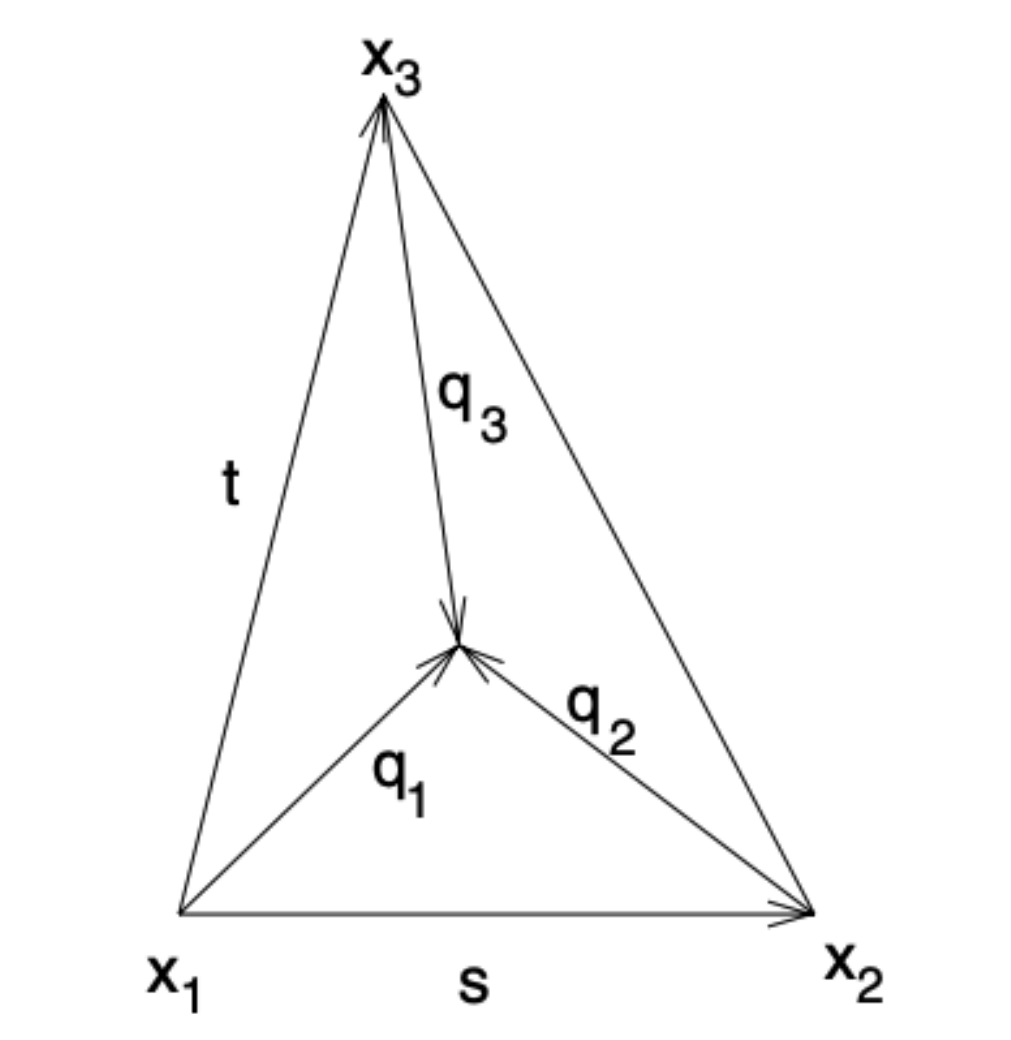}
	\includegraphics[width=\columnwidth]{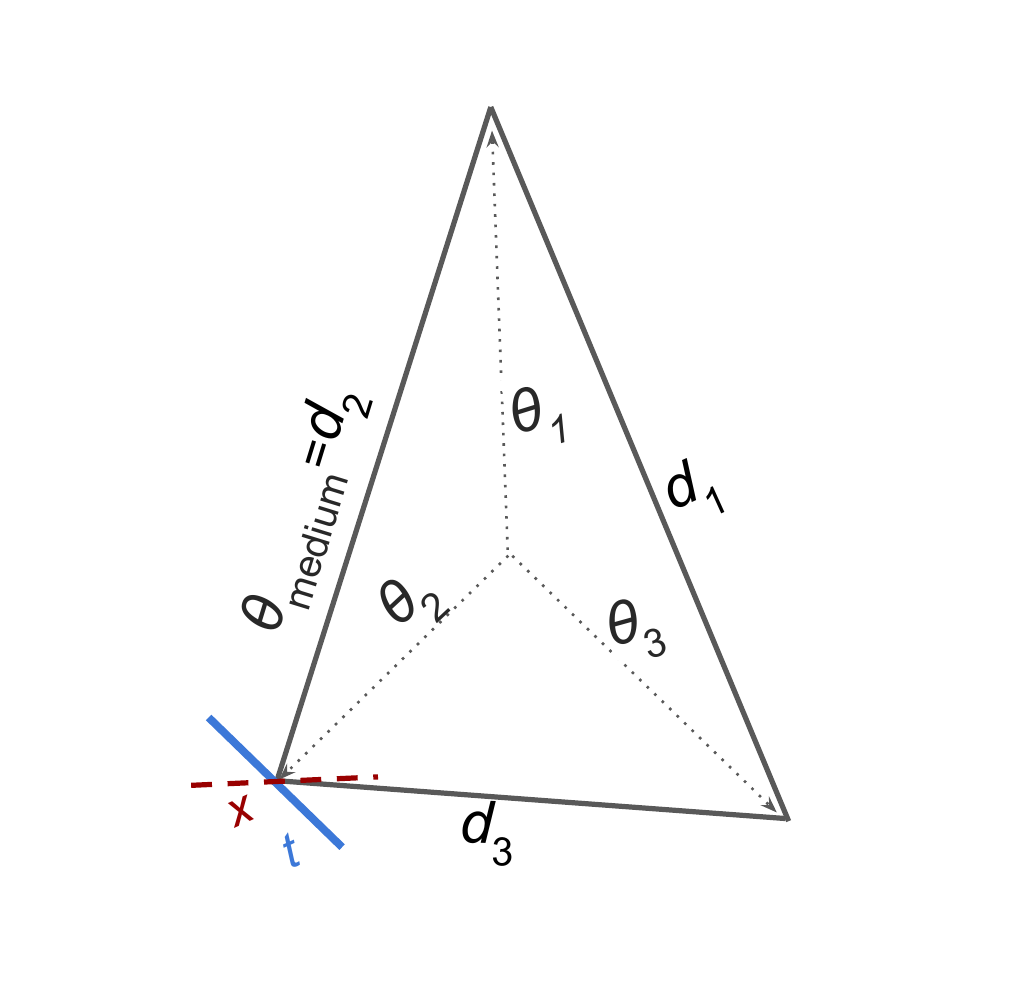}
    \caption{Convention for coordinates systems. \textbf{Top panel:} \textsc{TreeCorr} internal variables and side length definitions as utilized in eqs. (\ref{eq: q s t definition})-(\ref{eq: v definition}), reproduced from \cite{JBJ04}. \textbf{Bottom panel:} definition of distances of interest and shear projections ($t,\times$) relevant in the $\Gamma_i(\theta_\textrm{medium})$ and $\left\langle M_\textrm{ap} (\theta_1, \theta_2, \theta_3) \right\rangle$  measurements. Notably, the reference scale $\theta_\textrm{medium}$ is a side length, while the scales $\theta_i$ ($i=1,2,3$) are  radii from the triangle center.
    \label{fig: triangle notations}
    }
\end{figure}

Here we clarify some of the notation utilized in this draft, mainly in what refers to coordinates and definitions of triangle sides and their respective angles. We use the same conventions of \cite{JBJ04} and reproduce their Fig. 1 below in our Fig \ref{fig: triangle notations}.

With $\boldsymbol{q}_1$, $\boldsymbol{q}_2$ and $\boldsymbol{q}_3$ defined as the vectors from each of the triangle vertices to the centroid of the triangle, and $s$ and $\boldsymbol{t}'$ the sides of the triangle (notice that without loss of generality we fix $s$ and take $\boldsymbol{t}'$ to be at an angle $\alpha$ with respect to that line), we have 
\begin{equation}\label{eq: q s t definition}
\boldsymbol{q}_{1}=\frac{(s+\boldsymbol{t}')}{3},\quad\boldsymbol{q}_{2}=\frac{(\boldsymbol{t}'-2s)}{3},\quad\boldsymbol{q}_{3}=\frac{(s-2\boldsymbol{t}')}{3}.
\end{equation}

The functions $T_0$ and $T_1$ that enter the mass aperture computations such as eq.(\ref{eq: M3 integral}) are purely geometrical and dependent on the vectors above:
\begin{equation}\label{eq: T0 definition}
   T_{0}\left(s,\boldsymbol{t}\right)=-\frac{(\boldsymbol{q}_{1}^{*}\boldsymbol{q}_{2}^{*}\boldsymbol{q}_{3}^{*})^{2}}{24}\exp\left(-\frac{q_{1}^{2}+q_{2}^{2}+q_{3}^{2}}{2}\right)
\end{equation}
\begin{align}\label{eq: T1 definition}
T_{1}\left(s,\boldsymbol{t}\right) & =-\left(\frac{(\boldsymbol{q}_{1}\boldsymbol{q}_{2}^{*}\boldsymbol{q}_{3}^{*})^{2}}{24}-\frac{q_{1}^{2}\boldsymbol{q}_{2}^{*}\boldsymbol{q}_{3}^{*}}{9}+\frac{\boldsymbol{q}_{1}^{*2}+2\boldsymbol{q}_{2}^{*}\boldsymbol{q}_{3}^{*}}{27}\right)\nonumber\\
 & \times\exp\left(-\frac{q_{1}^{2}+q_{2}^{2}+q_{3}^{2}}{2}\right),
\end{align}
where bold symbols are vectors in complex notation with $x$/$y$ on the real/imaginary direction, eg $\boldsymbol{v}=v_x+iv_y$.

Additionally, internal \textsc{TreeCorr} units utilized to bin triangles are such that, for triangles of side lengths $d_1\leq d_2 \leq d_3$, we have 
\begin{equation}\label{eq: u definition}
    u=\frac{d_3}{d_2},
\end{equation}
\begin{equation}\label{eq: v definition}
    v=\pm \frac{(d_{1}-d_{2})}{d_{3}},
\end{equation}
where the positive and negative signs of $v$ correspond to whether side lengths are in clockwise or counterclockwise order respectively, and recall that we have named $\theta_\mathrm{medium}=d_2$ to conveniently bin the $\Gamma_i$ functions in eq.(\ref{eq: medium}). Note that with these definitions we have $u\in[0,1]$ and $v\in[-1,1]$. In practice, selecting i.e. equilateral triangles of characteristic side length $\theta_\mathrm{medium}$ within the output corresponds to sub-selecting the galaxies in bins $u\sim0$ and $v\sim0$.

\section{Signal-to-noise of a Vector}
\label{snr_vector}
The signal-to-noise ratio of a scalar value, $X$, with a Gaussian uncertainty, $\sigma$, is well-defined.
The signal is the expectation value of the measurement $\left\langle X\right\rangle$, and the noise is the standard deviation of the uncertainty $E$.
Thus, the signal-to-noise is simply the ratio of these.
\begin{align}
X &= \langle X \rangle + E \\
E &\sim \mathcal{N}(0, \sigma) \\
S/N(X)\ &\equiv \frac{ \langle X \rangle}{\sigma } \\
 &= \frac{\langle X \rangle} {\sqrt{\mathrm{Var}(X)}} \label{snr_scalar}
\end{align}
However, it is less obvious what the corresponding quantity should be for a vector $\mathbf{d}$, where each component of the vector
is itself a measurement with an uncertainty.  
We start by considering a data vector of independent measurements, each with its own Gaussian uncertainty.
\begin{align}
\mathbf{d} &= \{ d_i \} \\
d_i &= \langle d_i \rangle + E_i \\
E_i &\sim \mathcal{N}(0, \sigma_i)
\end{align}
We consider all possible linear combinations of the vector elements,
\begin{align}
X_\mathbf{w} &\equiv \mathbf{w} \cdot \mathbf{d} = \sum_i w_i d_i,
\end{align}
for arbitrary weight vectors $\mathbf{w}$.  For each choice of $\mathbf{w}$, the scalar quantity $X_\mathbf{w}$ of course has a well-defined signal-to-noise, given by Equation~\ref{snr_scalar},
but each choice may be different, depending on the specific weights being used.  
Among all such possible choices, we take the one with the largest signal-to-noise to define the signal-to-noise of the vector $\mathbf{d}$.
\begin{align}
S/N(\mathbf{d}) &\equiv \max_{\mathbf{w}} \left( \frac{\langle X_\mathbf{w} \rangle}{\sqrt{\mathrm{Var}(X_\mathbf{w})}}\right) \label{snr_def1}
\end{align}
We therefore need to determine what choice of weights $\mathbf{w}$ gives the largest signal-to-noise for $X_\mathbf{w}$.
For a given choice of $\mathbf{w}$, we have
\begin{align}
(S/N)^2 = \frac{ \left( \sum_j w_j \langle d_j \rangle \right)^2 }{ \sum_j w_j^2 \sigma_j^2 }.
\end{align}
As usual, we find $w_i$ at the extremum by setting the derivative to 0.
\begin{align}
0 = \frac{\partial \left(S/N\right)^2}{\partial w_i} &= \frac{2 \left(\sum_j w_j \langle d_j \rangle \right) \langle d_i \rangle}{\sum_j w_j^2 \sigma_j^2 }
- \frac{2 w_i \sigma_i^2 \left(\sum_j w_j \langle d_j \rangle \right)^2}{\left(\sum_j w_j^2 \sigma_j^2\right)^2} \nonumber \\
\langle d_i \rangle \sum_j w_j^2 \sigma_j^2 &= w_i \sigma_i^2 \sum_j w_j \langle d_j \rangle \nonumber \\
w_i &= \frac{\langle d_i \rangle}{\sigma_i^2}
\end{align}
The signal-to-noise for this choice of $\mathbf{w}$ is then
\begin{align}
S/N &= \frac{\sum_i w_i \langle d_i \rangle}{\sqrt{\sum_i w_i^2 \sigma_i^2}} \nonumber \\
&= \frac{\sum_i \langle d_i\rangle^2/\sigma_i^2}{\sqrt{\sum_i \left(\langle d_i \rangle/\sigma_i^2\right)^2 \sigma_i^2}} \nonumber \\
&= \sqrt{\sum_i \frac{\langle d_i \rangle^2} {\sigma_i^2}}.
\end{align}

Aside from the expectation value in the numerator, this is equivalent to $\sqrt{\chi^2}$, which is a relatively common approximation
used to estimate the signal-to-noise of a vector.  
Calculating the expectation value of $\chi^2$, we find
\begin{align}
\langle \chi^2 \rangle &= \left \langle \sum_i \frac{d_i^2}{\sigma_i^2} \right \rangle \nonumber \\
&= \sum_i \frac{\left( \langle d_i \rangle + E_i \right)^2}{\sigma_i^2} \nonumber \\
&= \sum_i \frac{\langle d_i \rangle^2 + 2 \langle d_i \rangle \langle E_i \rangle + \left\langle E_i^2 \right\rangle}{\sigma_i^2} \nonumber \\
&= \sum_i \frac{\langle d_i \rangle^2 + \sigma_i^2}{\sigma_i^2} \nonumber \\
&= (S/N)^2 + N_\mathrm{d.o.f.}.
\end{align}
Thus, we have derived the relatively simple relationship,
\begin{align}
S/N &= \sqrt{\langle \chi^2 \rangle - N_\mathrm{d.o.f.}} \label{snr_def2} .
\end{align}
In practice, one does not have access to the expectation value $\langle \chi^2 \rangle$, 
so we replace it with its measured value, which is the best we can do:
\begin{align}
S/N &= \sqrt{\chi^2 - N_\mathrm{d.o.f.}} \label{snr_def}.
\end{align}
For high signal-to-noise vectors, the approximation $S/N=\sqrt{\chi^2}$ is not bad.  
But when $\chi^2$ is only moderately larger than the number of degrees of freedom, the correction is important, 
and one should instead use Equation~\ref{snr_def}.
And of course if the measured $\chi^2$ is less than $N_\mathrm{d.o.f.}$, there is no detection, and the signal-to-noise is essentially zero.

Finally, what if the uncertainties are correlated?  That is, what if the data vector has a non-diagonal covariance matrix $C$?
It turns out that this case can be reduced to the same formula as above by diagonalizing $C$ and changing to
the basis where the covariance is diagonal.
\begin{align}
\mathrm{Cov}(\mathbf{d}) &\equiv C = V \Lambda V^T \\
\mathbf{z} &\equiv V^T \mathbf{d} \\
\mathrm{Cov}(\mathbf{z}) &= V^T \mathrm{Cov}(\mathbf{d}) V \nonumber \\
&= V^T V \Lambda V^T V \nonumber \\
&= \Lambda
\end{align}
Given our definition (Equation~\ref{snr_def1}), the signal-to-noise of $\mathbf{z}$ is the same as the signal-to-noise of $\mathbf{d}$.  
Furthermore, the $\chi^2$ for the two vectors are also equal:
\begin{align}
\chi^2 &= \mathbf{z}^T \Lambda^{-1} \mathbf{z} = \sum_i \frac{z_i^2}{\Lambda_{ii}} \nonumber \\
&= (V^T \mathbf{d})^T \Lambda^{-1} V^T \mathbf{d} \nonumber \\
&= \mathbf{d}^T V \Lambda^{-1} V^T \mathbf{d} \nonumber \\
&= \mathbf{d}^T (V \Lambda V^T)^{-1} \mathbf{d} \nonumber \\
&= \mathbf{d}^T C^{-1} \mathbf{d},
\end{align}
where we used the fact that $V^T = V^{-1}$.

We know that the signal-to-noise of $\mathbf{z}$
is given by Equation~\ref{snr_def}, since it has uncorrelated uncertainties.  
Since $\mathbf{d}$ has the same signal-to-noise as $\mathbf{z}$, and it has the same $\chi^2$ and $N_\mathrm{d.o.f.}$,
this must also be the correct formula for $\mathbf{d}$.  
Therefore, Equation~\ref{snr_def} applies even to a vector with a non-trivial covariance matrix.


\bibliographystyle{mnras_2author}
\bibliography{shear3pt, y3kp}

\end{document}